\documentclass[11pt,final]{article} 
\usepackage{authblk}
\usepackage[
backend=biber,
style=numeric,
sorting=nyt,
maxbibnames=999
]{biblatex}
\usepackage{enumitem}
\usepackage{fullpage}

\usepackage{amsmath}
\usepackage{amsthm}
\usepackage{amssymb}
\usepackage{thmtools, mathtools, thm-restate}

\usepackage[font=small]{caption}
\usepackage[final]{microtype}
\usepackage{subcaption}
\usepackage{xcolor}
\usepackage{tikz}
\usetikzlibrary{positioning}
\usetikzlibrary{trees}
\usetikzlibrary{decorations.pathreplacing}
\usepackage{pgfplots}
\pgfplotsset{compat=1.18}
\usepgfplotslibrary{dateplot}
\usepackage[english]{babel}
\usepackage{hyperref}
\usepackage[capitalise]{cleveref}
\usepackage[textsize=tiny]{todonotes} 
\usepackage{xspace}
\newtheorem{theorem}{Theorem}[section]
\newtheorem{lemma}[theorem]{Lemma}

\newtheorem{claim}[theorem]{Claim}
\newtheorem{property}[theorem]{Property}
\newtheorem{observation}[theorem]{Observation}
\newtheorem{proposition}[theorem]{Proposition}
\newtheorem{corollary}[theorem]{Corollary}
\newtheorem{definition}[theorem]{Definition}

\theoremstyle{remark}
\newtheorem{remark}[theorem]{Remark}
\usepackage[toc,page]{appendix}
\usepackage{algorithm}
\usepackage{algpseudocode}
\usepackage[appendix=append, bibliography=common]{apxproof}
\theoremstyle{plain}
\newtheoremrep{claimapp}[theorem]{Claim}
\crefname{claimapp}{Claim}{Claims}
\Crefname{claimapp}{Claim}{Claims}
\theoremstyle{plain}
\newtheoremrep{theoremapp}[theorem]{Theorem}
\crefname{theoremapp}{Theorem}{Theorems}
\Crefname{theoremapp}{Theorem}{Theorems}
\theoremstyle{plain}
\newtheoremrep{lemmaapp}[theorem]{Lemma}
\crefname{lemmaapp}{Lemma}{Lemmas}
\Crefname{lemmaapp}{Lemma}{Lemmas}
\algnewcommand{\Input}{\item[\textbf{Input:}]}
\algnewcommand{\Upon}{\item[\textbf{Upon}]}
\algnewcommand{\Atevery}{\item[\textbf{At every moment $t$:}]}
\algnewcommand{\for}{\item[\textbf{for}]}
\algnewcommand{\do}{\item[\textbf{do}]}
\algnewcommand{\if}{\item[\textbf{if}]}
\algnewcommand{\while}{\item[\textbf{while}]}
\algnewcommand{\then}{\item[\textbf{then}]}
\algnewcommand\algorithmicpriority{\textbf{Priority rule:}}
\algnewcommand\PRIORITY{\item[\algorithmicpriority]}

\usepackage{todonotes}

\newcommand{\comment}[1]{}

\newcommand{\SOL}{\mathsf{SOL}}

\newcommand{\allreqs}{\mathcal{R}} %
\newcommand{\ALG}{\mathsf{ALG}}

\newcommand{\OPT}{\widehat{\mathsf{OPT}}}
\newcommand{\md}{\mathcal{D}}
\newcommand{\mP}{\mathcal{P}}
\newcommand{\hl}{h^{\mathrm{left}}}
\newcommand{\hr}{h^{\mathrm{right}}}
\newcommand{\pr}{\mathcal{D}^\mathrm{p}}

\newcommand{\lamf}{\mathcal{F}} %
\newcommand{\Yhat}{\widehat{Y}}
\newcommand{\ql}{q^\mathrm{left}}

\newcommand{\Ir}{I^\mathrm{right}}
\newcommand{\Il}{I^\mathrm{left}}
\newcommand{\Jl}{J^\mathrm{left}}
\newcommand{\Jr}{J^\mathrm{right}}
\newcommand{\cross}{X}
\newcommand{\splitr}{S}
\newcommand{\mPo}{\mathcal{Q}}
\newcommand{\node}{\mathfrak{n}}
\newcommand{\Node}{\mathcal{N}}
\newcommand{\ver}{u}

\title{Noncrossing is All You Need: Beyond Identical Delay Functions in Online Matching with Delays on General Metrics}
\author{
Yossi Azar\thanks{School of Computer Science and AI, Tel-Aviv University, Tel-Aviv, Israel. Email: \mbox{azar@tauex.tau.ac.il}. Supported in part by the Israel Science Foundation grant No. 1728/26.}\hspace{2cm}
Junhao Gan\thanks{School of Computing and Information Systems, The University of Melbourne, Australia. Email: \mbox{junhao.gan@unimelb.edu.au}. Supported in part by the Australian Government through the Australian
Research Council ARC DP230102908.}\hspace{2cm}
Shahar Lewkowicz \thanks{School of Computer Science and AI, Tel-Aviv University, Tel-Aviv, Israel. Email: \mbox{shaharlewko22@gmail.com}.}
\and
Xiao Sun\thanks{School of Computing and Information Systems, The University of Melbourne, Australia. Email: \mbox{xiao.sun.1@student.unimelb.edu.au}. Supported by the Australian Government through the Australian Research Council
DP240101353 and by the University of Melbourne through the Melbourne Research Scholarship.}\hspace{2cm}
Seeun William Umboh\thanks{School of Computing and Information Systems, The University of Melbourne, Australia. ARC Training Centre in Optimisation Technologies, Integrated Methodologies, and Applications (OPTIMA), Melbourne, Australia. Email: \mbox{william.umboh@unimelb.edu.au}. Supported by the Australian Government through the Australian Research
Council DP240101353.}
}

\date{}

\begin{document}
\begin{titlepage}
\def\thepage{}
\thispagestyle{empty}
\maketitle
\setcounter{page}{0}
\begin{abstract}
Online optimization with delays has emerged as a powerful paradigm with numerous applications, including matching, caching, server management, network design, and set cover. In most of these problems, competitive algorithms are known for arbitrary monotone delay functions, even when different requests are associated with different delay functions. The notable exception is online matching with delays, where all previously known algorithms achieving constant or polylogarithmic competitive ratios require the delay functions to be identical, either linear or concave, and known in advance.

We study the online minimum-cost perfect matching with delays problem on general metrics. 
We introduce the broad class of noncrossing delay functions, which strictly generalizes the classic setting of identical concave delay functions while naturally modeling scenarios in which different requests may experience different delay costs. We present the first deterministic $O(1)$-competitive algorithm for the single-location online matching problem with noncrossing delay functions in the non-clairvoyant setting as well as a randomized $O(\log n\log \Delta + \log^2 \Delta)$-competitive algorithm of an arbitrary metric space where $n$ is the number of points of the metric space and $\Delta$ is its aspect ratio. Our result allows each request to have an arbitrary noncrossing delay function, thereby simultaneously removing the assumptions of identical delay functions, concavity, and clairvoyance that were required by all previous constant or polylogarithmic-competitive algorithms.

\end{abstract}
\end{titlepage}
\clearpage

\section{Introduction}
\label{sec:intro}
Over the last decade,
online optimization with delays has emerged as a powerful and widely applicable paradigm. It has been successfully applied to a broad range of problems, including matching, caching, server management, network design, set cover, data structures, and many others (e.g., \cite{bienkowski2016online, Online_matching_haste_makes_waste, AzarGGP2017, azar2020setcoverd, DBLP:conf/focs/AzarT20, DBLP:conf/soda/DufayW26}, among many others). In this framework, requests arrive online and may be served at any time after their arrival. Delaying service incurs a request-dependent penalty that increases over time, while postponing decisions may reduce the underlying optimization cost by allowing the algorithm to 
aggregate requests and serve them in a batch.
Consequently, the objective is to decide both which requests to serve and when to serve them so as to minimize the sum of the traditional optimization cost and the accumulated delay penalties. This fundamental trade-off has made the framework both practically relevant and a rich source of algorithmic ideas in the theory of online algorithms.

A remarkable feature shared by nearly all of these problems is that competitive algorithms are known for arbitrary monotone delay functions, even when different requests are associated with different delay functions.
The main exception to this phenomenon is the matching problem, first studied by~\textcite{Online_matching_haste_makes_waste}, which was also the first problem studied in the online-with-delays framework. All previously known algorithms achieving constant or polylogarithmic competitive ratios require the delay functions to be identical and either linear or concave. In this paper, we break this barrier by presenting, for the first time, competitive algorithms for a broad family of non-identical delay functions.

We note that one possible explanation for why matching is fundamentally different from other online optimization problems with delays is the lack of input monotonicity. For virtually all other problems studied in this framework, adding more requests can only increase (or leave unchanged) the value of the optimal solution. In contrast, for matching, the arrival of additional requests may arbitrarily decrease the optimal cost by creating new matching opportunities.
This fundamental non-monotonicity appears to be a major obstacle to extending techniques developed for other online optimization problems with delays to the matching problem.

One motivation for the matching with delays comes from commercial platforms for matching clients or players. Examples include gaming platforms such as chess.com and bridgebase.com, where players arrive online with the intention of participating in a game against another player. Since most games involve exactly two participants, the platform must match arriving players into pairs. While players are generally willing to tolerate some waiting time before being matched, excessive delays reduce user satisfaction. This naturally raises the following algorithmic question: \emph{how should the platform match players so as to minimize the overall delay cost}?

Beyond gaming platforms, this problem captures a wide range of online matching applications, including roommate assignment, carpooling, and other real-time matching services. 

\paragraph{Problem Definition:} In this paper, we study the online minimum-cost perfect matching with delay problem, introduced by~\textcite{Online_matching_haste_makes_waste}. Formally, requests arrive online over time in a metric space $(X,d)$. At any point after their arrival, the algorithm may match any pair of unmatched requests or postpone their matching until a later time.
 
The cost of a matching is determined by how requests are matched into pairs and when they are matched. Each request $r$ arrives at location $\ell(r) \in X$, has an arrival time $a(r)$, and a delay function $f_r(t)$ that starts at 0 at $a(r)$, and is nondecreasing with time. When $r$ is matched at time $m(r)$, it incurs a delay cost of $f_r(m(r))$. 
If request $r$ is matched to request $r'$, they incur a connection cost of $d(\ell(r),\ell(r'))$.
The total cost of a matching is %
\begin{equation*}
\sum_r f_r(m(r))+\sum_{(r,r')\in M}d(\ell(r),\ell(r')),
\end{equation*}
where $M$ is the matching and the objective is to minimize this quantity.

\paragraph{Previous delay setting:} Previous works obtaining constant or polylogarithmic competitive ratios rely simultaneously on the following three assumptions: 
\begin{enumerate}[noitemsep]
 \item 
all requests share the \emph{same delay function} (shifted by the arrival time)\footnote{More specifically, $f_r(a(r) + d) = f_q(a(q) + d)$ for any two requests $q,r$ and $d \geq 0$.};
 \item 
the delay function is \emph{linear} or \emph{concave};
 \item 
the delay function is known upon the request's arrival (i.e., the algorithm is \emph{clairvoyant}). 
\end{enumerate}

The MPMD problem was introduced by~\textcite{Online_matching_haste_makes_waste}, who studied the case of linear delay functions. Later, the authors in~\cite{Impatient_Online_Matching} considered convex delay functions and posed the question of whether comparable competitive guarantees could be obtained for concave delay functions. This question was subsequently answered by~\textcite{DBLP:conf/soda/AzarRV21}, who developed competitive algorithms for the concave setting.

\paragraph{Our delay setting:} In contrast, we study a substantially more general setting in which
\begin{enumerate}[noitemsep]
\item 
each request may have a \emph{different delay function};
\item 
delay functions are \emph{arbitrary noncrossing} functions (defined formally in \cref{sec:prob-def});
\item 
the algorithm is \emph{non-clairvoyant}, namely, for every request that has already arrived, its delay function is revealed only up to the current time, and its future behavior is unknown.
\end{enumerate}

Informally, noncrossing delay functions form a sequence of functions in which, at every point in time, an earlier request always has a delay value at least as large as that of any later request, while its derivative is never larger. Observe that noncrossing delay functions generalize the classic setting of identical concave delay functions studied by~\textcite{DBLP:conf/soda/AzarRV21}.
Noncrossing delay functions naturally arise in practice. Earlier clients are typically more frustrated than later arrivals, and their frustration may evolve in an arbitrary manner depending on external circumstances, rather than following a pre-defined concave shape. Nevertheless, it is reasonable to assume that, at any point in time, an earlier client is never less frustrated than a later one, while the rate at which this frustration increases is never higher than that of a later client.

\paragraph{Our Contributions:}
In this paper, we consider both the single location case and an arbitrary metric space. %
We present the first deterministic \(O(1)\)-competitive algorithm for the single-location online matching problem with noncrossing delay functions in the nonclairvoyant setting. For general metrics with \(n\) points and aspect ratio \(\Delta\), we obtain a randomized $O(\log n\log \Delta + \log^2 \Delta)$-competitive algorithm. Our algorithms allow arbitrary request-specific delay functions subject only to the noncrossing condition, thereby simultaneously removing the assumptions of identical delay functions, concavity, and clairvoyance imposed by previous constant or polylogarithmic-competitive algorithms.

\begin{theorem}\label{constant-competitive}
For any metric space with \(n\) points and aspect ratio \(\Delta\), there is a randomized nonclairvoyant $O(\log n\log \Delta + \log^2 \Delta)$-competitive algorithm for the online matching problem with noncrossing delay functions. For the special case of a single location, there is a deterministic nonclairvoyant \(O(1)\)-competitive algorithm.
\end{theorem}

We also show that it is impossible for any deterministic clairvoyant algorithm to achieve a bounded competitive ratio without the assumption that an earlier request always has a delay cost at least as large as that of any later request, even for the single point case. %
See \cref{prop:noncrossing-necessary} in \cref{sec:noncrossing-necessary}.

\begin{toappendix}
\label{sec:noncrossing-necessary}
\begin{proposition}
    \label{prop:noncrossing-necessary}
    There is no deterministic clairvoyant algorithm with bounded competitive ratio if later requests are allowed to have larger delay cost than earlier requests.
\end{proposition}
\begin{proof}
Consider the following adversarial instance where all requests arrive at the same location. Initially, two requests $r_1$ and $r_2$ arrive simultaneously at time $a(r_1)=a(r_2)=0$, and both incur delay that grows linearly at unit rate. If the algorithm never matches them, then no additional requests arrive, and its competitive ratio is unbounded.
Otherwise, suppose the algorithm matches $r_1$ and $r_2$ at time $t$. At time $t+1$, a new request $r_3$ arrives whose delay grows linearly at rate $\Delta$. Finally, request $r_4$ arrives at time $2t+2$. The optimal solution matches $r_3$ with $r_1$ and $r_4$ with $r_2$, incurring a total cost of $3t+3$, whereas the algorithm incurs a cost of at least $\Delta(t+1)$. Therefore, the competitive ratio is at least $\Delta/3$ which is unbounded as $\Delta$ can be arbitrarily large.
\end{proof}
\end{toappendix}

\paragraph{Geometric Interpretation for single location:} Our problem for the very special case of single location admits the following equivalent geometric interpretation. Points are born over time at the origin and continuously move to the right along the $x$-axis, where the $x$-coordinate of a point represents its accumulated delay. The relative order of the points is preserved over time (by the noncrossing property), and the distance between any two consecutive points—and hence between any pair of points—never increases. At any time, the algorithm may freeze any pair of points, remove them from the system, and incur a cost equal to the sum of their current positions, corresponding to matching the two associated requests. Every point must eventually be frozen. The objective is to minimize the total freezing cost, which is exactly the total delay incurred before all requests are matched.

Despite this simple and intuitive formulation, designing a constant-competitive online algorithm, and proving its competitiveness, turns out to be remarkably challenging.
An obvious first approach is to match the requests greedily (i.e., delay only if there is no other request).  While this yields an optimal matching when delay functions are identical and linear (or convex), %
unfortunately, as observed by \cite{DBLP:conf/soda/AzarRV21}, the simple greedy matching algorithm %
has unbounded competitive ratio even for identical concave delay functions. To see this, consider an instance in which $2m$ requests arrive at times $\{0\} \cup \{i, i+\varepsilon\}_{i = 1}^{m-1} \cup \{m\}$ and the delay function of each request $r$ is given by 
$f_r(t)=\min\{t-a(r), 1\}$.  
The greedy algorithm incurs a total cost of $1+(1-\varepsilon)(m-1)$. In contrast, the optimal solution matches the first request with the last and matches all remaining requests greedily, thereby yielding a total cost of $1+(m-1)\varepsilon$. By choosing $\varepsilon$ sufficiently small, the competitive ratio of the greedy algorithm becomes at least $m$, and is therefore unbounded. This example illustrates that achieving a bounded competitive ratio requires sacrificing short-term cost in order to obtain greater future savings.

\paragraph{Arbitrary metric space versus single location.}
As explained above, even the single-location case, a highly restricted special case of a general metric space—requires a nontrivial algorithm. Surprisingly, our first technique is to “reduce” the general case to single-location problems. We use a hierarchically separated tree (HST) and run a single-location algorithm at each vertex, processing requests promoted from its children according to a carefully designed rule.

This is not a reduction in the usual sense, as it requires a stronger guarantee than competitiveness for the single-location problem. Specifically, at each vertex, we compare the cost of the online algorithm on the requests that reach that vertex, the ``virtual instance" (which filters out some of the requests), with the optimal cost for all requests originating in its subtree, the ``consolidated instance". Thus, the online algorithm and the offline optimum operate on different instances. Although this “pseudo-reduction” requires a more powerful single-location result, it significantly simplifies the analysis of the general metric case. The main challenge then lies in designing and analyzing a single-location algorithm that admits this stronger comparison between the virtual instance and consolidated instances, though even the standard single-location problem is not substantially simpler.
This “pseudo-reduction” incurs a multiplicative loss proportional to the height of the HST, which is \(O(\log \Delta)\), together with an additional \(O(\log \Delta)\) loss in the delay-cost term alone. Embedding the original metric space into HSTs introduces an \(O(\log n)\) distortion in the connection-cost term. Since the latter two losses affect separate cost components, they combine additively. Accounting for the initial height-dependent loss, the resulting competitive ratio is \(O(\log \Delta(\log n+\log \Delta))=O(\log n\log \Delta+\log^2\Delta)\).

\subsection{Our Techniques}
\label{sec:techniques}
As we already discussed the pseudo-reduction from an arbitrary metric space to virtual single location instances, most of this subsection deals with an actual single location instance which includes most of the new ideas and techniques. At the end we mention the additional difference that we encounter while switching from to comparing the virtual instance (which filters out some of the requests) to the consolidate instance for generalizing our algorithm from a single location to an arbitrary metric space.

Our work builds on the technique of matching based on counters that increase over time, which is central to previous works on online matching with delays, and develops new insight on this technique. %
The main challenge is in designing the matching condition so that the following desiderata are satisfied: (1) the total counter value captures the cost of the algorithm and (2) is also not much larger than the cost of the optimal solution. Intuitively, an equivalent view is that the algorithm matches when the cost of the match can be charged to some set of counters which can then be charged to the optimal solution.

As our problem, even for single location, generalizes the single-location problem of \cite{DBLP:conf/soda/AzarRV21}, it is useful to contrast our approach with theirs. Our algorithm differs significantly from theirs, which relies heavily on the fact that all delay functions are identical. In particular, that algorithm maintains counters that are shared among all requests. In contrast, in our setting it is natural to associate a separate counter with each request. Whenever a request's counter increases, it does so at the same rate as its delay function.

It is not difficult to see that it suffices to consider algorithms that match only the latest two pending requests (only for the single location case), since %
later requests always have delay derivatives at least as large as those of earlier requests. To ensure that the matching condition is eventually satisfied for the last two pending requests, our algorithm increases only the counter of the latest pending request. %

Unfortunately, as demonstrated by our first two attempts, the counter values alone do not provide sufficient information to determine when a match should be performed. This is in sharp contrast to previous algorithms, where the counters alone were sufficient. To overcome this difficulty, we introduce a carefully designed potential function based on a partition of the requests, as described later.

\paragraph{First attempt.}
Motivated by the counterexample to the greedy algorithm, a natural first attempt is to wait until the latest request's counter reaches half of the second latest request to make a match. %
It also has the benefit that all pending requests' counter values form a geometric sequence at every moment, and thus the total delay seems to be bounded by the biggest term, i.e. the counter value of the earliest request. Unfortunately, it does not work under  uniform linear delays. Intuitively, it fails because it matches requests too slowly while the previous requests have already accumulated huge cost. In particular, its cost can be much larger than the total counter value. See \cref{sec:counterexamples} for a specific counterexample.

\begin{toappendix}
\section{Counterexamples}
\label{sec:counterexamples}
\paragraph{Counterexample of the first attempt.}
The first attempted algorithm increases the counter of the latest pending request with its instantaneous increase rate and matches the latest two requests $(r,r')$ if the counter value of $r'$ reaches half of that of $r$.

We construct a sequence of instances $\{I_k\}$, and then show that the competitive ratio of the algorithm is unbounded as $k$ tends to infinity. The delay function is the uniform linear delay that $f_r(t)=t-a(r)$ for every request $r$ and $t\geq a(r)$. For every $I_k$, we use $\ALG_k$ to denote the cost of the algorithm, $\ell_k$ to denote the time difference between the arrival of the first request and the time all requests are matched by the algorithm, and $z_k$ to denote the final counter value of the first request when it is matched. Under the uniform linear delay, it is easy to see that $\ell_k$ is an upper bound of the optimal cost, and we will compare $\ALG_k$ against $\ell_k$ to give a lower bound of the competitive ratio. 

The first instance $I_1$ has two requests in total: $a(r_1)=0, a(r_2)=1$. Under the algorithm, $r_1$ and $r_2$ are matched at time $3/2$. As a result, we have $\ALG_1=2, \ell_1=3/2$ and $z_1=1$.

We now show how to construct $I_{k+1}$ when $I_k$ is defined. The first request of $I_{k+1}$ arrives at time $0$. At time $2z_k$, we start to repeat a copy of $I_k$, until the moment when all requests in this copy of $I_k$ are matched, and immediately at the same moment the first request of the next copy of $I_k$ arrives, and so on. We repeat $I_k$ for $N_k$ times in this way, where $N_k$ is a large integer. The moment when all requests of the $N_k$'th repetition are matched by the algorithm, the last request of $I_{k+1}$ arrives.

\begin{figure}[h]
\centering
\begin{tikzpicture}[scale=0.80,
    dot/.style={circle,fill,inner sep=1.3pt},
    every node/.style={font=\small}
]

\def\L{2.4}      %
\def\ZL{2.2}     %
\def\ZR{1.8}     %
\def\H{1.3}      %
\def\BH{2.8}     %

\coordinate (A) at (0,0);
\coordinate (B) at (\ZL,0);
\coordinate (C) at ({\ZL+\L},0);
\coordinate (D) at ({\ZL+2*\L},0);
\coordinate (E) at ({\ZL+4*\L},0);   %
\coordinate (F) at ({\ZL+5*\L},0);
\coordinate (G) at ({\ZL+5*\L+\ZR},0);

\draw (A)--(G);

\foreach \p in {A,B,C,D,E,F}
    \node[dot] at (\p) {};

\draw (B) to[out=70,in=110] node[midway,above] {$I_k$} (C);
\draw (C) to[out=70,in=110] node[midway,above] {$I_k$} (D);
\draw (E) to[out=70,in=110] node[midway,above] {$I_k$} (F);

\draw (A) to[out=70,in=110]
    node[midway,above] {$I_{k+1}$}
    (G);

\node at ({\ZL+2.9*\L},0.42) {$\cdots\cdots$};

\draw[decorate,decoration={brace,mirror,amplitude=5pt}]
    (A) -- node[below=6pt] {$2z_k$} (B);

\draw[decorate,decoration={brace,mirror,amplitude=5pt}]
    (B) -- node[below=6pt] {$\ell_k$} (C);

\draw[decorate,decoration={brace,mirror,amplitude=5pt}]
    (C) -- node[below=6pt] {$\ell_k$} (D);

\draw[decorate,decoration={brace,mirror,amplitude=5pt}]
    (E) -- node[below=6pt] {$\ell_k$} (F);

\draw[decorate,decoration={brace,mirror,amplitude=5pt}]
    (F) -- node[below=6pt] {$z_k$} (G);

\end{tikzpicture}
\caption{An illustration of the construction of $I_{k+1}$ with $I_k$, where we use dots on the axis to represent the first and last requests of $I_{k+1}$ as well as the first request of each copy of $I_k$. The arcs represent the time period of each $I_k$ and $I_{k+1}$ (from the arrival of the first request until all requests in it have been matched) rather than matches.}
\end{figure}

With the inductive argument, it is easy to see that the following observations hold, when we are allowed to choose the way of breaking ties.

\begin{observation}
For every $k\geq 1$:
    \begin{itemize}
        \item The algorithm always matches the first request with the last request in $I_k$.
        \item $z_{k+1}=2z_k$ and hence $z_k=2^{k-1}$.
        \item $\ell_{k+1}=N_{k}\cdot\ell_k+3z_k$ and $\ALG_{k+1}=N_k\cdot\ALG_k+N_k\cdot\ell_k+4z_k$.
    \end{itemize}
\end{observation}

Now we consider the ratio between $\ALG_{k+1}$ and $\ell_{k+1}$: given $\ALG_{k}$, $\ell_{k}$ and $z_{k}=2^{k-1}$, the ratio between $\ALG_{k+1}$ and $\ell_{k+1}$ tends to $\ALG_k/\ell_k+1$ when $N_k$ tends to infinity. Therefore, we can choose appropriate values of $N_k$'s such that $\ALG_{k+1}/\ell_{k+1}\geq \ALG_k/\ell_k+1/2$ for every $k\geq 1$, and so the algorithm has unbounded competitive ratio.
\end{toappendix}

\paragraph{Second attempt.}
The first attempt can be seen in the following way: the second latest request $r$'s counter value is saturated from below when the latest request's counter reaches half of its value. It fails because of the ``memorylessness" that saturates too slowly. A natural way around this is to consider not only the latest request at a given moment, but also all previous requests that used to be the next pending request after $r$ (we will use ``children" of $r$ to denote them in \cref{sec:online-laminar}) and all their counters contribute to the saturation of $r$'s counter.

Therefore, our second attempt is to match when the sum of counter values of all children of $r$ reaches half of the second latest request $r$'s counter value. Under this algorithm, the counter value of each request is within a constant factor of its actual incurred delay, so the issue with the counterexample of the first attempt is addressed. Unfortunately, this algorithm also fails because it might match ``too fast" instead, and the total counter value can be much larger than the cost of the optimal solution. We present the counterexample in \cref{sec:counterexamples}. In this example, the requests' delays have similar patterns to concave delays: initially increasing fast but stopping at some value; while we note that they are not uniform concave delays and the delay function of a given request may not be concave either.

\begin{toappendix}
\paragraph{Counterexample of the second attempt.}
The second attempted algorithm increases the counter of the latest pending request with its instantaneous increase rate and matches the latest two requests $(r,r')$ when the sum of counter values of all children (see \cref{sec:online-laminar} for the definition of children, as well as roots that we are going to use later) of $r$ reaches half of the counter value of $r$ itself.

We still construct a sequence of instances $\{I_k\}$. $I_{1}$ comprises a sufficiently long sequence of requests where the time differences between two consecutive arrivals are $1,1/2,1,1/2,\ldots,1$ chronologically, and the delay function of each request $r$ is $f_r(t)=\min \{t-a(r),1\}$. The algorithm always matches two consecutive requests at time $1/2$ after the arrival of the later one in the pair. Let $N_1$ be the number of pairs in $I_1$ and $A_1$ be the sum of counter values of all roots of the laminar forest induced by the algorithm on $I_1$, then $A_1=N_1$. One feasible solution is to match the first request with the last request and all other consecutive requests from early to late into pairs, which provides an upper of the optimal cost $\OPT_1=\frac{1}{2}(N_1+1)$. While slightly abusing the notation, we will use $\OPT_k$ to denote the cost of the suggested solution for every $I_k$, which provides an upper bound of the optimal cost.

Now we show how to construct $I_{k+1}$ recursively with $I_k$. We first state the arrival times of all requests and then we define the delay function associated with each request. Let $A_k$ be the sum of final counter values of all roots of the laminar forest induced by the algorithm on $I_k$, which will be used as a lower bound of the algorithm's cost. The first request of $I_{k+1}$ arrives at time $0$. At time $2A_k+2$, we start a copy of $I_k$, until the moment when all requests in this copy of $I_k$ are matched by the algorithm, then a new request arrives exactly at this moment. Time $1$ after the arrival of the last request in the above \emph{process}, we start to repeat this process. Let $N_{k+1}$ be the total number of such processes (including the first process).

\begin{figure}[h]
\centering
\begin{tikzpicture}[scale=0.80,
    dot/.style={circle,fill,inner sep=1.3pt},
    every node/.style={font=\small}
]

\def\L{2.4}      %
\def\ZL{2.0}     %
\def\ZR{1.4}     %
\def\H{1.3}      %
\def\BH{2.8}     %

\coordinate (A) at (0,0);
\coordinate (B) at (\L,0);
\coordinate (C) at ({\L+\ZL},0);
\coordinate (D) at ({\L+\ZL+\ZR},0);
\coordinate (E) at ({2*\L+\ZL+\ZR},0);   %
\coordinate (F) at ({3*\L+\ZL+\ZR},0);
\coordinate (G) at ({3*\L+2*\ZL+\ZR},0);
\coordinate (H) at ({3*\L+2*\ZL+2*\ZR},0);

\draw (A)--(H);

\foreach \p in {A,B,C,D,E,F,G}
    \node[dot] at (\p) {};

\draw (B) to[out=70,in=110] node[midway,above] {$I_k$} (C);
\draw (F) to[out=70,in=110] node[midway,above] {$I_k$} (G);

\draw (A) to[out=70,in=110]
    node[midway,above] {$I_{k+1}$}
    (H);

\node at ({\L+1.6*\ZL+\ZR},0.42) {$\cdots\cdots$};

\draw[decorate,decoration={brace,mirror,amplitude=5pt}]
    (A) -- node[below=6pt] {$2A_k+2$} (B);

\draw[decorate,decoration={brace,mirror,amplitude=5pt}]
    (C) -- node[below=6pt] {$1$} (D);

\draw[decorate,decoration={brace,mirror,amplitude=5pt}]
    (E) -- node[below=6pt] {$2A_k+2$} (F);

\draw[decorate,decoration={brace,mirror,amplitude=5pt}]
    (G) -- node[below=6pt] {$1$} (H);
\draw[decorate,decoration={brace,mirror,amplitude=5pt}]
    (0,-0.9)--(5.8,-0.9)
    node[midway,yshift=-0.45cm]{process};
\draw[decorate,decoration={brace,mirror,amplitude=5pt}]
    (8.2,-0.9)--(14.0,-0.9)
    node[midway,yshift=-0.45cm]{process};
\end{tikzpicture}
\caption{An illustration of the construction of $I_{k+1}$ with $I_k$, where we use dots on the axis to represent the first and last requests of each process as well as the first request of each copy of $I_k$, and the request that arrives at the end of a process is the first request of the following process. The arcs represent the time period of each $I_k$ and $I_{k+1}$ (from the arrival of the first request until all requests in it have been matched) rather than matches.}
\end{figure}

We now describe the delay function of each request in the first repetition of such process, and the delays of requests in other repetitions are essentially the same functions shifted by their arrival times. The first request $r$'s delay is $f_r(t)=\min\{t-a(r),2A_k+2\}$. The last request of the process has the same delay increase shifted by its arrival time. For each request in a copy of $I_k$: their delay increases in the same way as the corresponding request in $I_k$ (shifted by arrival times) until the end when all requests in $I_k$ are matched by the algorithm, then it starts to increase at the same rate as the next request that arrives after it (the maximum possible increase rate while maintaining the noncrossing delays), until its total delay reaches $2A_k+2$ and then it stops increasing hereinafter. Such delays satisfy the noncrossing property.

We have the following observations when we are allowed to choose the way of breaking ties for the algorithm:

\begin{observation}
    For every $k\geq1$:
    \begin{itemize}
        \item The first request in each process has all roots under the implementation of the algorithm on $I_k$ and the last request of the process as all its children, and the algorithm always matches the first and last requests in the same process.
        \item $A_{k+1}=2(A_k+1)N_{k+1}$.
        \item $\OPT_{k+1}=N_{k+1}(\OPT_k+1)+2A_k+1$ by matching the first and last requests of $I_{k+1}$, all requests in the same copy of $I_k$ in the same way as $\OPT_k$, and the last request of a process with the first request of the following process.
    \end{itemize}
\end{observation}

Given $A_k$ and $\OPT_k$, the ratio between $A_{k+1}$ and $\OPT_{k+1}$ tends to $2(A_k+1)/(\OPT_k+1)$ when $N_{k+1}$ is sufficiently large, which tends to $2A_k/\OPT_k$ by letting $N_k$ (and thus $\OPT_k$) tend to infinity. Therefore, we can choose appropriate values of $N_k$'s to make $A_{k+1}/\OPT_{k+1}\geq 3/2\cdot A_k/\OPT_k$ for every $k\geq 1$, and so the algorithm has unbounded competitive ratio.

\end{toappendix}

In summary, both attempts fail to match requests at an appropriate speed under some cases, including linear delays that grow fast and other delays that grow slowly. This also illustrates that our noncrossing delay model comprises a broad range of ways that delays can evolve over time. The first attempt shows that the algorithm cannot be memoryless, and the second attempt shows that such memory cannot be the simple sum of all children's counters of the second latest request.

\paragraph{Partition-based potential.}
At a high level, our algorithm assigns a potential to each pending request based on the counter values of its current children. The potential is designed so that, at any point in time, only one child can increase the potential of its parent. The algorithm matches a pair of requests once the potential of the earlier request reaches half of its counter value. This perspective also unifies the two previously discussed failed attempts: they correspond to defining the potential as either the counter of a single child or the sum of the counters of all current children. Roughly speaking, the potential measures the amount of delay incurred by the optimal solution that can be safely charged by our algorithm. The potential function and the matching rule are carefully designed so that a request is matched precisely when its accumulated counter value can be fully charged to the optimal solution without causing any overcharging.

The key ingredient in our charging argument is a decomposition of the requests that arrive after a given request according to the intervals used by the optimal solution. We consider the maximal intervals of the optimal solution that overlap the lifetime of the request without completely covering it. These intervals partition the requests that arrive while the request is pending. Unfortunately, since the optimal solution is unknown, the corresponding partition is also unknown. Instead of relying on this partition directly, we define the potential function to represent the value of the minimum-value partition.

Now consider one such interval that contains several consecutive children of the request but not the request itself. Since the interval begins before the first of these children arrives and remains active until after all but the last have been matched, the noncrossing property implies that the delay accumulated by the request during this period is no larger than the delay incurred by that interval in the optimal solution. This observation enables us to charge the delay accumulated over each such period to the corresponding interval of the optimal solution without overcharging.

\paragraph{Handling virtual instances for HSTs.}

When extending our single-location algorithm to HSTs, we construct a virtual instance at each vertex by filtering out certain requests arriving in its subtree. Counter updates at an internal vertex must also account for the parity of the number of pending requests in its subtree, a quantity also used in earlier counter-based algorithms on trees (\cite{DBLP:conf/soda/AzarCK17,DBLP:conf/soda/AzarRV21}). Accordingly, the virtual algorithm at each vertex has an odd state and an even state, determined by this parity, which specifies which counter to increase.

A key difference from the single-location algorithm discussed earlier is that we do not necessarily increase the counter of the latest request. As a result, the property that the algorithm always matches the latest two requests no longer holds, and the laminarity of the solution may fail. To address this issue, we sometimes terminate a request's current interval and counter and create new ones corresponding to the same request. This allows us to partition the pending time of each request into multiple intervals, with the resulting collection of intervals being laminar. In this way, we can adapt the proof techniques developed for the single-location case to those laminar intervals and their associated counters.

\subsection{Related Work} 
The MPMD problem with linear delay functions was introduced by~\textcite{Online_matching_haste_makes_waste}, who presented a randomized algorithm with competitive ratio $O(\log^2 n+\log \Delta)$, where $\Delta$ denotes the aspect ratio of the underlying metric. Later,~\textcite{DBLP:conf/soda/AzarCK17} improved this bound to $O(\log n)$, thereby eliminating the dependence on $\Delta$. They also proved an $\Omega(\sqrt{\log n})$ lower bound on the competitive ratio of any randomized algorithm. This lower bound was subsequently strengthened by the authors in~\cite{DBLP:conf/approx/AshlagiACCGKMWW17} to $\Omega(\log n/\log\log n)$, nearly closing the gap. The upper bound of $O(\log n)$ was later generalized for identical concave delay functions by \cite{DBLP:conf/soda/AzarRV21}.

The authors in~\cite{Impatient_Online_Matching} later studied MPMD with convex delay functions on $n$-point uniform metrics. They first established an $\Omega(n)$ lower bound on the competitive ratio of any deterministic algorithm. Their lower-bound construction uses delay functions of the form $f(t)=t^\alpha$, for $\alpha>1$. They also presented a deterministic $O(n)$-competitive algorithm under the same class of delay functions. \textcite{DBLP:conf/approx/GanSU26a} complemented this study by presenting a deterministic $O(1)$-competitive algorithm for polynomial delays such that $f'(0)>0$.

The only non-clairvoyant delay models that have been studied in prior works are set and size-based delays proposed by Deryckere and Umboh~\cite{DBLP:conf/approx/DeryckereU23}. They showed a deterministic lower bound of $\Omega(\Delta)$ for set delays and gave a randomized upper bound of $O(m^4)$ for size-based delays, where $m$ is the number of requests. The results on size-based delays were subsequently improved by~\textcite{DBLP:conf/approx/GanSU26a} and they showed that the randomized competitive ratio is $\Omega(n)$ and $O(n^2)$ for every metric, refuting the possibility of a polylogarithmic competitive ratio.

A related line of work considers Minimum-Cost Bipartite Perfect Matching with Delays (MBPMD), the bipartite variant of MPMD with linear delay functions, in which requests can only be matched to requests of the opposite polarity. The authors in~\cite{DBLP:conf/approx/AshlagiACCGKMWW17} established an $\Omega(\sqrt{\log n/\log\log n})$ lower bound on the competitive ratio of any randomized algorithm. They also presented two $O(\log n)$-competitive randomized algorithms: the first extends the algorithm of~\textcite{Online_matching_haste_makes_waste} to the bipartite setting, while the second extends the algorithm of~\textcite{DBLP:conf/soda/AzarCK17}.

The deterministic variants of MPMD and MBPMD with linear delay functions have also attracted considerable attention. For deterministic algorithms and further developments, see \cite{A_Match_in_Time_Saves_Nine:_Deterministic_Online_Matching_with_Delays, A_Primal_Dual_Online_Deterministic_Algorithm_for_Matching_with_Delays, AzarF2018, DBLP:conf/soda/DufayW26, ahmadian2026generalframeworkmatchingdelays, EmekSW2019}.

Although this paper focuses on online matching with delays, the notion of delays has been extensively studied in many other online optimization problems. Such models have numerous applications spanning operations management, operating systems, supply chain management, and other domains. Examples include the online service with delays problem \cite{AzarGGP2017, BienkowskiKS2018, azar2019general}, the multi-level aggregation problem \cite{bienkowski2016online, buchbinder2017depth, azar2019general}, and many others. For a broader overview of online optimization problems with delays, we refer the reader to \cite{carrasco2018online, gupta2020caching, DBLP:journals/disopt/Epstein21, The_Price_of_Clustering_in_Bin-Packing_with_Applications_to_Bin_Packing_with_Delays, azar2020setcoverd, DBLP:conf/focs/AzarT20,DBLP:conf/isaac/Touitou21,DBLP:conf/icalp/ChenKU22,DBLP:conf/stoc/Touitou23}.

Finally, we note that in other optimization problems with delays,   there has been much recent interest in studying the power of clairvoyance \cite{azar2020setcoverd,DBLP:conf/soda/LeUX23,Touitou23,DBLP:conf/stacs/Touitou25,DBLP:journals/algorithmica/EzraLPRU26,DBLP:conf/icalp/DinitzFU26} and novel delay models \cite{moseley2025putting,holdingandbacklogAL,DBLP:conf/soda/ShmoysSU26,DBLP:conf/approx/BhorePU26}.%

\paragraph{Paper Organization:} We start in \cref{sec:prob-def} by formally defining the problem and introducing some useful notations and basic properties, including laminar solutions and randomized embedding of general metrics into $\sigma$-HSTs. We introduce the pseudo-reduction to single-location instances in \cref{sec:pseudo-reduction}, and present our algorithm on a $2$-HST of height $H$ by describing the algorithms that run on each of its vertices in \cref{sec: laminar and our algorithm}. We prove that our global algorithm is $(O(H), O(H^2))$-competitive on the $2$-HST in \cref{sec:analysis}, thus completing the proof of \cref{constant-competitive}. We conclude with some remarks and open problems in \cref{sec: conclusions}.

\section{Preliminaries}

\label{sec:prob-def}

\subsection{Problem Definition and Notation}
In the online matching with delays problem, a set $\allreqs$ of requests arrive over time (we assume WLOG that no two requests arrive at the same time) in a given metric space $(X,d)$, and each request incurs delay cost until it is matched with another request. More precisely, each request $r$ has a \emph{location} $\ell(r)\in X$, and a continuous \emph{delay} function $f_r(t)$ that is initially 0 at the arrival time of $r$, which we denote by $a(r)$, and is nondecreasing with time. In addition, we assume that the algorithm is nonclairvoyant: at every moment $t$, the algorithm receives $f'_r(t)$, the instantaneous delay increase rate of $r$, for every request $a(r)\leq t$ that has arrived.

A solution to the problem consists of a pair $(M, m)$ where $M$ is a matching, and $m(r)$ is the time when request $r$ is matched, for each request $r$. When requests $r,r'$ are matched, we denote the match as $(r,r')$ writing the requests in order of arrival, i.e., $r$ arrives earlier.
When request $r$ is matched at time $m(r)$, the total delay incurred is $f_r(m(r))$. The goal is to match the requests so as to minimize the total delay of requests (\emph{delay cost}) plus the distance between locations of every matched pair (\emph{connection cost}), i.e., $\sum_{r \in \allreqs} f_r(m(r))+\sum_{(r,r')\in M}d(\ell(r),\ell(r'))$.

For $I \subseteq \mathbb{R}^+$, we write $f_r(I) = \int_{t \in I} f'_r(t) dt$ and $f_r(m(r)) = \int_{t \in [a(r),m(r))} f'_r(t) dt$. Throughout the paper, for a set of requests $S$, we write ``$i$-th latest request of $S$" to refer to the request in $S$ with the $i$-th latest arrival time. For any two requests $r_1,r_2$, we write $r_1 \preceq r_2$ if $a(r_1) \leq a(r_2)$, and $r_1 \prec r_2$ if $a(r_1) < a(r_2)$.
\begin{definition}
    [Noncrossing delays]
    \label{def:noncrossing}
    An instance $\allreqs$ has \emph{noncrossing delays} if for every $r_1\prec r_2$, it holds that $f_{r_{1}}'(t)\leq f'_{r_{2}}(t)$ and $f_{r_{1}}(t)\geq f_{r_{2}}(t)$ at every time $t\geq a(r_{2})$.
\end{definition}
Noncrossing delays means that the delay of every request increases at least as fast as the delay of an earlier-arriving request, but its total delay is no larger than the total delay of the earlier-arriving request. Equivalently, for any two requests, the difference in their total delays is always nonincreasing and nonnegative.
\subsection{Laminar Solutions}
\label{sec:def-laminar}
In this part, we introduce the concept of laminarity, which is an important property that holds for every optimal solution on a single location instance (where the connection cost is always zero). Furthermore, the laminar property of a family of intervals will be used in defining our algorithm on metrics as well. 
\begin{definition}
    [Request intervals and laminarity]
    \label{def:laminar}
    Let $\mathcal{I}$ be a single-point instance where requests' delays are non-crossing and $(M,m)$ be a solution of $\mathcal{I}$. For each request $r$, define its \emph{interval} to be $I_r = [a(r),m(r))$. The solution is a \emph{laminar solution} if the set of  intervals $I_r$ is laminar.
\end{definition}
The following is immediate from the definition.
\begin{proposition}
\label{prop:lam}
    A solution $(M,m)$ is laminar if and only if for every request $r$ and every request $r'$ with arrival time $a(r') \in (a(r),m(r))$, $m(r')\leq m(r)$.
\end{proposition}
From \cref{prop:lam}, it is easy to see that an online algorithm that only matches the latest two pending requests is laminar. Moreover, since there is an optimal solution that only matches the latest two pending requests on a single-location instance (because their delays increase at the fastest rate), and the optimal solution always matches a pair of requests as soon as they have both arrived to avoid additional delay, we have the following lemma. %
\begin{lemma}\label{non-crossing}
    For online matching with noncrossing delays on a single location, there is an optimal solution $(M^*, m^*)$ that is laminar, and every match $(r,r') \in M^*$ is matched on arrival of the later request $r'$.
\end{lemma}
Although the laminarity of the optimal solution does not hold on metric spaces beyond the single location, we will use \cref{non-crossing} in the charging scheme for our algorithm analysis.
\subsection{Reduction to $2$-HSTs}
We solve the online matching with noncrossing delay on general metrics by reducing it to $2$-HSTs. This is the key technique that has been used for all previous results of the online matching with delay problem on general metrics, e.g.~\cite{Online_matching_haste_makes_waste,DBLP:conf/soda/AzarCK17,DBLP:conf/soda/AzarRV21}. We introduce the definition of a $(\beta,\gamma)$-competitive algorithm to separate delay and connection costs.
\begin{definition}
    For an arbitrary instance of online matching with noncrossing delays, let $\ALG$ be a randomized online algorithm and its cost and $\SOL$ be an arbitrary solution. Furthermore, $\SOL_d$ and $\SOL_t$ denote the connection and delay costs incurred by $\SOL$, respectively. We say that $\ALG$ is $(\beta,\gamma)$-competitive if $\mathbb{E}[\ALG]\leq \beta\cdot\SOL_d+\gamma\cdot\SOL_t$.
\end{definition}
The reason that we focus on solving the problem on $\sigma$-HSTs is because we can randomly embed general metrics into them. We first formally define the notion of Hierarchically Separated Trees (HSTs) and embedding a metric into another.
\begin{definition}[$\sigma$-HST]
A $\sigma$-hierarchically well-separated tree ($\sigma$-HST) is a rooted tree such that: all edges at a particular level have the same weight and all leaves are at the same level (combinatorial depth), and edge lengths on every root-to-leaf path decrease by a factor of $\sigma$ at each step.
\end{definition}
Note that in our definition, we require that every leaf has the same combinatorial depth. This key requirement is the reason that we cannot use weighted HSTs to reduce the height as per the techniques that have been used for linear and concave delays \cite{DBLP:conf/soda/AzarCK17,DBLP:conf/soda/AzarRV21}.

\begin{definition}[Metric embeddings]
Let $\mathcal{M} = (X,d)$ be a finite metric
space, and let $\mathcal{D}$ be a probability distribution over metrics on a finite set $X'$. We say that $\mathcal{M}$ embeds into
$\mathcal{D}$ if $X \subseteq X'$, and for every $x,y \in X$ and every metric space $\mathcal{M}' = (X',d')$ in the support of $\mathcal{D}$, we have
$d(x,y) \leq d'(x,y)$. The distortion of this embedding is
defined to be
\begin{equation*}
    \max_{x\neq y\in X}\frac{\mathbb{E}_{\mathcal{M}' = (X',d')\sim \mathcal{D}}[d'(x,y)]}{d(x,y)}.
\end{equation*}
\end{definition}

We use the following result from \cite{DBLP:journals/jcss/FakcharoenpholRT04} to embed any general $n$-point metric with aspect ratio $\Delta$ into a distribution over $2$-HSTs.

\begin{lemma}
\label{lem: randomized embedding}
    Any $n$-point metric can be embedded, with distortion $O(\log n)$, into a distribution supported on metrics induced by $2$-HSTs with height $O(\log\Delta)$.
\end{lemma}

The next lemma introduced in \cite{Online_matching_haste_makes_waste} shows how to convert a matching’s cost from a metric in the support of the embedding to that on the original metric.

\begin{lemma}
\label{lem: cost conversion between metrics}
Suppose that a metric $\mathcal{M}$ can be embedded into a distribution $\mathcal{D}$ with distortion $\mu$. Additionally suppose that for every metric space in the support
of $\mathcal{D}$ exists an algorithm that is $(\beta, \gamma)$-competitive for the MPMD problem.
Then, there exists a $(\mu\beta, \gamma)$-competitive algorithm for the MPMD problem on the
metric $\mathcal{M}$.
\end{lemma}

In remaining parts of the paper, we describe the algorithm on a $2$-HST whose height is $H$ (including the root level), denoted by $\mathcal{T}=(V,E)$. For every vertex $\ver\in V$, let $L_\ver$ be the distance between $\ver$ and its leaf descendant, $T_\ver$ be the subtree rooted at $\ver$, and $p(\ver)$ be its parent. From the definition of $2$-HSTs and properties of geometric sequences, we have $L_{p(\ver)}\geq 2L_\ver$ as long as $\ver$ is not the root of $\mathcal{T}$. If $\ver$ is the root, then define $L_{p(u)}=\infty$. Moreover, we use $\allreqs_\ver$ to express the set of all requests that arrive at leaves of $T_\ver$. For an interval $I \subseteq \mathbb{R}^+$, we define $R_\ver(I) = \{r \in \allreqs_\ver \mid a(r) \in I \}$, i.e., the set of requests that arrive at leaves of $T_\ver$ during $I$.

\section{A Pseudo-Reduction to Single-Location Instances}
\label{sec:pseudo-reduction}

In \cref{sec: reduction to singles}, we reduce the description of our algorithm $\ALG$ on $\mathcal{T}$ to its subroutines $\ALG_\ver$ for each vertex $\ver$ of the HST, including the requests that arrive in each instance $\mathcal{I}_\ver$, and how we translate $\ALG_\ver$ to $\ALG$ while losing a constant factor only (\cref{lemma: reduction to single}). In \cref{sec: consolidated}, we introduce the consolidated instance induced within each $T_\ver$ (\cref{consolidated instance}) and prove their relations with any global solution $\SOL$ (\cref{lemm: levels of SOL}), which will be used in proving the competitive ratio of $\ALG$. We abuse the notations by using $\ALG$ and $\ALG_\ver$ to refer to both the solution and its cost, the same for $\SOL$ and $\OPT_\ver$ (defined in~\cref{sec: consolidated}).

\subsection{Transform Virtual Algorithms $\ALG_\ver$'s to $\ALG$}
\label{sec: reduction to singles}
\paragraph{Algorithm overview.} Our algorithm $\ALG$ creates several single-location instances $\mathcal{I}_u$ for each vertex $u$ of the HST; these are called \emph{virtual instances}. The requests in the virtual instances are called \emph{virtual copies} of actual requests. The arrivals of virtual copies in virtual instances are based on a global mechanism where $\ALG$ promotes every request to different levels of its ancestors based on its total incurred delay. We then run a single-location algorithm on each of these instances independently. When two virtual copies of requests are matched, the global algorithm decides whether or not to match the corresponding actual requests. We will prove that the total cost of $\ALG$ is within a constant factor of that of all $\ALG_\ver$'s:

\begin{lemma}[Reduction to Single-Location]\label{lemma: reduction to single}
$\ALG\leq 4\sum_{\ver\in V}\ALG_\ver$.
\end{lemma}

\paragraph{Virtual instances.} 
At every moment $t$, we say that a pending request $r$'s \emph{position} is $\ver$ if $\ver$ is the ancestor of its leaf location $\ell(r)$ (or $\ell(r)$ itself) such that $L_\ver\leq f_r(t)<L_{p(\ver)}$. Each request's position is uniquely defined at every moment when it is unmatched by $\ALG$ (note that the position of $r$ is different from its \emph{location} $\ell(r)$, where the latter is its arrival location in the metric).

We now describe the single-location virtual instance $\mathcal{I}_\ver$ for each $\ver\in V$. For each request $r$ that arrives at $\ell(r)$ and vertex $\ver$ that is an ancestor of $\ell(r)$, a virtual copy of it arrives in $\mathcal{I}_\ver$ the moment $t$ when $f_{r}(t)=L_\ver$. Or equivalently, its arrival time in $\mathcal{I}_\ver$ is the earliest moment when $r$'s position is $\ver$ ($r$ is \emph{promoted} to $\ver$). Since each $r$ creates at most one such virtual copy in any $\mathcal{I}_\ver$, we still use $r$ to denote this virtual copy without ambiguity.

After the arrival of virtual $r$ in $\mathcal{I}_\ver$, its delay increase in $\mathcal{I}_\ver$ is the delay increase of $r$ while its position is $\ver$. We emphasize that, later on, when the position of the virtual $r$ is an ancestor of $\ver$, the virtual copy of $r$ \textbf{remains} in $\mathcal{I}_\ver$ and its delay in $\mathcal{I}_\ver$ stops increasing after $f_r(t)$ reaches $L_{p(\ver)}$. In summary, the delay function $f_{\ver,r}(t)$ of a request $r$ that arrives in $\mathcal{I}_\ver$ at time $f_r^{-1}(L_\ver)$ (without loss of generality, assume that $f_r'(t)>0$ for every $r$ and $t\geq a(r)$, then $f_r(t)$ is a bijection) can be expressed as:
\[
    f_{\ver,r}(t) = \begin{cases}
        f_r(t)-L_\ver & \text{when $L_\ver\leq f_r(t)<L_{p(\ver)}$;} \\
        L_{p(\ver)}-L_\ver & \text{when $f_r(t)\geq L_{p(\ver)}$.}
    \end{cases}
    \]

In other words, $\mathcal{I}_\ver$ receives part of request $r$'s delay increase while its total value is between $L_\ver$ and $L_{p(\ver)}$. It is easy to see that the delays $f_{\ver,r}(t)$ for virtual copies of $r$ in $\mathcal{I}_\ver$ satisfy the non-crossing properties as well, and this is why we treat each $\mathcal{I}_\ver$ as a virtual instance. We note that those virtual instances of different vertices are not independent of each other, because the set of requests that are promoted to $\ver$ depends on the output of other instances whose vertices are in the subtree $T_\ver$: $\mathcal{I}_\ver$ only receives requests that have not been matched by $\ALG$ when their positions are descendants of $\ver$, and we can regard $\ALG$'s matches of requests positioned at descendants of $\ver$ as a filter removing some requests from $\mathcal{I}_\ver$. 

We observe that when $r$ is positioned at $\ver$, its delay increase rate in $\mathcal{I}_\ver$ is $f'_{\ver,r}(t)=f'_r(t)$, otherwise $f'_{\ver,r}(t)=0$ if $r$ has been promoted to a higher level. The next observation is immediate from the mechanism of $\ALG$'s promotion of requests to different levels based on their total delays and the symmetric structure of a HST.

\begin{observation}
\label{obs: promotion}
For every $\ver\in V$, the arrival order of requests in $\mathcal{I}_\ver$ is the same order as in the actual instance. Moreover, if $r_1$'s position is $\ver_1$ and $r_2$'s position is $\ver_2$ such that $\ver_2$ is a descendant of $\ver_1$, then $r_1\prec r_2$.
\end{observation}

We will use it in our charging method for analysis.

\paragraph{Virtual to actual matches.} Now we describe how to transform the virtual matches produced by $\ALG_\ver$'s to the actual solution output by $\ALG$. For every vertex $\ver$, if $\ALG_\ver$ matches two requests $r_1$ and $r_2$, then there are two cases according to whether the position of each of $r_1$ and $r_2$ is $\ver$ at the moment. If $r_1$ and $r_2$ both are still positioned at $\ver$, then $\ALG$ matches $r_1$ and $r_2$ the moment when they are matched by $\ALG_\ver$. If at least one of $r_1$ and $r_2$ has been promoted to a higher level, then $\ALG$ ``ignores'' the action of $\ALG_\ver$ matching them and both of them will be promoted to $p(\ver)$ when the total delay reaches $L_{p(\ver)}$, and how each of $r_1$ and $r_2$ is eventually matched by $\ALG$ will depend on other sub-routines at ancestors of $\ver$. As a result, the actual matches produced by $\ALG$ can be significantly different from those made by its sub-routines $\ALG_\ver$'s.

\begin{toappendix}
\begin{proof}[Proof of \cref{lemma: reduction to single}]
We first fix a vertex $\ver$ and consider the total delay incurred by each request $r$ when it is positioned at $\ver$, for both $\ALG_\ver$ and $\ALG$, and then compare them. There are two cases to consider. (Case 1) If $r$ stays unmatched in $\mathcal{I}_\ver$, then its delay while being positioned at $\ver$ is $L_{p(\ver)}-L_\ver$ for both $\ALG_\ver$ and $\ALG$. (Case 2) For every matched pair $(r,r')$ such that $r\prec r'$ by $\ALG_\ver$, we discuss two different subcases according to whether the position of each of them is $\ver$ when $\ALG_\ver$ makes the match. (Case 2a) If both are positioned at $\ver$, then $r$ and $r'$ are matched by $\ALG$ at the same time, and the actual delay of $r$ and $r'$ incurred when they are positioned at $\ver$ is equal to their virtual delay in $\mathcal{I}_\ver$. (Case 2b) If $r$'s position is not $\ver$, then the total delay incurred by $r$ when it is positioned at $\ver$ is $L_{p(\ver)}-L_\ver$ by both $\ALG$ and $\ALG_\ver$, which is a lower bound of the sum of the virtual delay of $r$ and $r'$ in $\mathcal{I}_\ver$. Since the actual delay of them incurred while being positioned at $\ver$ is no more than $2(L_{p(\ver)}-L_\ver)$, we lose a factor of at most $2$. By summing up all vertices of $\mathcal{T}$, we conclude the total delay cost of $\ALG$ is no more than $2\sum_{\ver\in V}\ALG_\ver$. 

Because $\ALG$ only matches a pair of requests with connection cost $2L_\ver$ when they have both been promoted to $\ver$, its connection cost is no more than the corresponding requests' total actual delays. As a result, the connection cost of $\ALG$ is no more than $2\sum_{\ver\in V}\ALG_\ver$ as well.

Therefore, the total cost of $\ALG$ is no more than $4\sum_{\ver\in V}\ALG_\ver$.
\end{proof}
\end{toappendix}

The proof of \cref{lemma: reduction to single} can be found in the Appendix. As a result, with $\ALG$'s global mechanism of promoting requests to different levels of their ancestors, it suffices to describe the sub-algorithm $\ALG_\ver$ for each vertex $\ver\in V$ in \cref{sec: laminar and our algorithm}.

\subsection{Consolidated Instances}
\label{sec: consolidated}
For every $\ver\in V$, we define a consolidated single-location instance comprising all requests whose locations are within $T_\ver$, the subtree rooted at $\ver$.

\begin{definition}[Consolidated instance induced within $T_\ver$]
\label{consolidated instance}
The input request set is the same as $\mathcal{R}_{\ver}$, all requests whose locations are leaves of $T_\ver$. For every request $r\in\mathcal{R}_{\ver}$, its arrival time in the consolidated instance is also $a(r)$ as in the original instance on $\mathcal{T}$. The total delay incurred by $r$ until $t$ if unmatched is equal to $\hat{f}_r(t)=\min\{f_r(t),L_{p(\ver)}\}$, i.e. it increases at the same rate as $f_r(t)$ until its value is capped by $L_{p(\ver)}$.
\end{definition}
It is easy to see that such delays are non-crossing. However, different from the conventional online matching with delays, here for each $\ver$ that is not the root of $\mathcal{T}$, the number of requests in the instance is not necessarily even, i.e. there might be requests unmatched at the end. Since the delay function of each request is capped by a constant $L_{p(\ver)}$, this is the cost of leaving one request unmatched, and so the objective is well-defined. We also note this consolidated instance is different from the virtual instance $\mathcal{I}_\ver$ that is described in~\cref{sec: reduction to singles}: here every request that arrives within $T_\ver$ is counted from its arrival time, rather than the moment when it is promoted to $\ver$, and the consolidated instance includes those not promoted to $\ver$ by $\ALG$ as well. This is why we call it a ``pseudo-reduction'', because we compare the cost of $\ALG_\ver$ against a different instance. For every request $r\in\mathcal{R}_\ver$, it holds that $\hat{f}_r(t)\geq f_{\ver,r}(t)$ and $\hat{f}'_r(t)= f'_{\ver,r}(t)$ for all $t\geq f_r^{-1}(L_\ver)$.

We use $\OPT_u$ to denote the optimal solution as well as its cost of the consolidated instance induced within $T_\ver$. Without loss of generality, we assume that $\OPT_\ver$ is laminar (\cref{non-crossing}), and it matches all requests into pairs if $|\mathcal{R}_\ver|$ is even and leaves exactly one request unmatched if $|\mathcal{R}_\ver|$ is odd.

Given a solution $\SOL$ of the online matching with noncrossing delay instance on $\mathcal{T}$, let $\SOL_d$ be its connection cost and $\SOL_t$ be its delay cost. Moreover, for every vertex $\ver$, we use $\SOL_{\ver,d}$ to denote its cost of using the edge connecting $\ver$ and $p(\ver)$ to make a match (zero if $\ver$ is the root) and $\SOL_{\ver,t}$ to denote part of its delay cost incurred by requests whose locations are within $T_\ver$. It holds that $\SOL_d=\sum_{\ver\in V}\SOL_{\ver,d}$ and $H\cdot\SOL_t=\sum_{\ver\in V}\SOL_{\ver,t}$, where the latter is because the delay incurred by each request is counted for $H$ times along the path from its leaf location to the root.

We now establish the relation between $\SOL$ and $\OPT_\ver$'s, whose proof we defer to the Appendix.
\begin{lemma}[Consolidated to Global]
\label{lemm: levels of SOL}
    For any solution $\SOL$, $\sum_{\ver\in V}\OPT_\ver\leq 2\SOL_{d}+H\cdot\SOL_{t}$.
\end{lemma}
\begin{toappendix}
\begin{proof}[Proof of \cref{lemm: levels of SOL}]
Because $\SOL_d=\sum_{\ver\in V}\SOL_{\ver,d}$ and $H\cdot\SOL_t=\sum_{\ver\in V}\SOL_{\ver,t}$, it suffices to show that $\OPT_\ver\leq 2\SOL_{\ver,d}+\SOL_{\ver,t}$ for every $\ver \in V$.

Given any $\SOL$, we construct a solution of the consolidated instance in the following way: if $\SOL$ matches a pair $(r,r')$ such that both $r$ and $r'$ are in $\mathcal{R}_\ver$, then we match them on arrival of the later request; if $\SOL$ matches $r\in\mathcal{R}_\ver$ with another request not in $\mathcal{R}_\ver$, then $r$ stays unmatched.

Now we compare the cost of this solution to $\SOL_{\ver,t}$ and $\SOL_{\ver,d}$. For every matched pair $(r,r')$, the delay cost of the earlier request is no more than that under $\SOL$ because of our definition of the capped delay, and this part of the delay is counted as part of $\SOL_{\ver,t}$. For every unmatched request $r$, $\SOL$ matches it with a request outside the subtree $T_\ver$, which incurs cost of $L_{p(\ver)}-L_{\ver}$ of using the edge connecting $\ver$ to $p(\ver)$. Because of the geometric property of $2$-HST, this is at least $\frac{1}{2}L_{p(\ver)}$, half the capped delay in the consolidated instance. As a result, $\OPT_\ver\leq 2\SOL_{\ver,d}+\SOL_{\ver,t}$.
\end{proof}
\end{toappendix}

With \cref{lemma: reduction to single,lemm: levels of SOL}, to prove the competitive ratio of $\ALG$, it suffices to compare the total cost of all $\ALG_\ver$'s against that of all $\OPT_\ver$'s. At the end of \cref{sec:analysis}, we will show that $\sum_{\ver \in V}\ALG_\ver\leq O(1)\sum_{\ver \in V}\OPT_\ver$ (\cref{charge alg to opt}).

In the remainder of the algorithm description and analysis, we may consider each vertex $\ver$ separately by fixing $\ver \in V$ and only considering request arrivals within $T_\ver$ without explicitly specifying it every single time.

\section{Single-Location Algorithms}
\label{sec: laminar and our algorithm}

 In this section, we define $\ALG_\ver$ for every $\ver$, which completes the description of $\ALG$ combined with the way to translate $\ALG_\ver$'s to $\ALG$ that has been defined in \cref{sec: reduction to singles}. In \cref{sec:online-laminar}, we use laminarity defined in \cref{sec:def-laminar} to define a forest structure of laminar intervals, which is key to applying tree terminologies including parents and children to the potential function and algorithm in \cref{sec:counters-matching}.

\subsection{Forest Structure of Laminar Intervals}
\label{sec:online-laminar}

In this subsection, we fix a vertex $\ver$ of $\mathcal{T}$, and define a forest structure $\lamf_\ver(t)$ that evolves over time, where each \emph{node} represents a time interval (we reserve ``vertices'' to mean vertices of $\mathcal{T}$ and ``nodes'' to mean those of the forest $\lamf_\ver(t)$). Moreover, the intervals in the forest form a laminar family, and the interval of each node $\node$ is a subset of the pending time of a request that arrives in $\mathcal{I}_u$, which we use $r(\node)$ to denote. We also call $r(\node)$ the \emph{request} of node $\node$. On the other hand, we note that the solution of the single-point instance $\mathcal{I}_\ver$ itself is not necessarily laminar as those intervals in the forest are not the same as the request pending intervals. 

We use $\Node_\ver$ to denote the set of all nodes that are added to $\lamf_\ver(t)$ at some time $t$. Once a node is added to $\lamf_\ver(t)$, it will never be deleted, while it can be \emph{terminated}. We use $a_\ver(\node)$ to denote the moment when node $\node$ is added to $\lamf_\ver(t)$ (we also call it the arrival time of $\node$), and $m_\ver(\node)$ to denote the moment when $\node$ is terminated, for which the conditions we will specify later. For each node $\node$ of $\lamf_\ver(t)$, we define the notion of online node intervals at moment $t$.

\begin{definition}
    [Online intervals]
    Given an online algorithm on $\mathcal{I}_\ver$, the \emph{interval $I_\node(t)$ at time $t$} of node $\node$ is defined as follows:
    \[
    I_\node(t) = \begin{cases}
        [a_\ver(\node), m_\ver(\node)) & \text{if $\node$ has already been terminated} \\
        [a_\ver(\node), t) & \text{otherwise}
    \end{cases}
    \]
\end{definition}

Same as the terminology that we use for requests, for a set of nodes $S$ in $\Node_\ver$, we write ``$i$-th latest node of $S$" to refer to the node in $S$ with the $i$-th latest arrival time.  For any two nodes $\node_1,\node_2\in\Node_\ver$, we write $\node_1 \preceq_\ver \node_2$ if $a_\ver(\node_1) \leq a_\ver(\node_2)$, and $\node_1 \prec_\ver \node_2$ if $a_\ver(\node_1) < a_\ver(\node_2)$.

Without specifying the exact rules of matching requests of $\ALG_\ver$, for this part we only need the following property to make the laminar forest structure well-defined:

\begin{property}
\label{property: semi-laminar}
At every moment, among all currently live nodes $\node_1\prec_\ver\ldots\prec_\ver\node_k$ and their requests $r(\node_1),\ldots,r(\node_k)$, the online algorithm only matches either $r(\node_{k-1})$ and $r(\node_{k})$, or $r(\node_{k-2})$ and $r(\node_{k-1})$.
\end{property}

\noindent We will see later in \cref{sec:counters-matching} that $\ALG_\ver$ indeed satisfies this property.

\paragraph{Update rules of $\lamf_\ver(t)$.} Now we define the rules of updating $\lamf_\ver(t)$, given any online algorithm of $\mathcal{I}_\ver$ that satisfies~\cref{property: semi-laminar}. There are two events that change the structure of $\lamf_\ver(t)$: when a new request arrives in $\mathcal{I}_\ver$, and when two requests are matched by the online algorithm. 

When a request $q$ arrives in $\mathcal{I}_\ver$, if there is no live node, then we create a new node $\node$ that is the root of a new tree in $\lamf_\ver(t)$, and let $r(\node)=q$. If $\node'$ is the latest live node right before $q$ arrives in $\mathcal{I}_\ver$, then we create a new node $\node$ as a child of $\node'$ and let $r(\node)=q$.

When two requests are matched, if they are the requests of the latest two live nodes in $\lamf_\ver(t)$ at the moment, then we terminate those two nodes; if they are the requests of the third and second latest live nodes, then we terminate those two nodes together with the latest live node, and create a new node as a child of the fourth latest live node (if it exists) or the root of a new laminar tree (otherwise), and let the request of this node be that of the previous latest live node that has just been terminated.

We can apply \cref{prop:lam} to the termination time of interval nodes in $\lamf_\ver(t)$ as well, which implies the following proposition.

\begin{proposition}
    \label{prop:online-lam}
    At every time $t$, the intervals $I_\node(t)$ form a laminar family.
\end{proposition}

We apply tree terminology to nodes as follows: node $\node$ is said to be a \emph{parent}, \emph{child}, \emph{ancestor}, \emph{descendant}, or \emph{sibling} of node $\node'$ at time $t$ if $I_\node(t)$ is a parent, child, ancestor, descendant, or sibling of $I_{\node'}(t)$ in $\lamf_\ver(t)$, respectively.  Denote by $C_\node(t)$ the children of $\node$ at time $t$. We also define the roots of the trees to be siblings of each other.  See \cref{fig:laminar} for an illustration of $\lamf_\ver(t)$.

\begin{figure}[ht]
\centering

\begin{subfigure}[t]{0.45\textwidth}
    \centering
\begin{tikzpicture}[scale=0.7]

\draw[thick, ->] (0,0) -- (10,0) node[right] {Time};
    
\foreach \x in {1,...,9}{
    \draw[thick] (\x,0) -- (\x,0); 
    \fill (\x,0) circle (2pt);
    \node[below] at (\x,0) {$\x$};
}

\draw[thick, dashed] (1,5) -- (9,5);
\fill (1,5) circle (2pt);

\draw[thick, dashed] (2,4) -- (9,4);
\fill (2,4) circle (2pt);

\draw[thick, dotted] (9,5) -- (9,0);

\draw[thick] (3,3) -- (4.5,3);
\fill (3,3) circle (2pt);
\draw[thick,dotted] (4.5,3) -- (4.5,0);

\draw[thick] (5,3) -- (8,3);
\fill (5,3) circle (2pt);
\draw[thick, dotted] (8,3) -- (8,0);

\draw[thick] (4,2) -- (4.5,2);
\fill (4,2) circle (2pt);

\draw[thick] (6,2) -- (8,2);
\fill (6,2) circle (2pt);
\draw[thick, dotted] (8,2) -- (8,0);

\draw[thick] (7,1) -- (8,1);
\fill (7,1) circle (2pt);

\draw[thick,dotted] (8,3) -- (9,3);
\fill (8,3) circle (2pt);
\end{tikzpicture}
\caption{The online intervals at time $t = 9$.}
\label{fig:intervals}
\end{subfigure}
~
\begin{subfigure}[t]{0.45\textwidth}
\centering
\begin{tikzpicture}[scale=0.7,
  every node/.style={
    circle,
    draw,
    minimum size=6mm,
    inner sep=1pt
  },
  level distance=13mm,
  sibling distance=28mm
]

\node {$I_1$}
  child {
    node {$I_2$}
      child { node {$I_3$} 
            child { node {$I_4$}  edge from parent[red, very thick]}}
      child {
        node {$I_5$}
          child {
    node {$I_6$}
    child {
        node[draw=black,thin, text=black] {$I_7$}
        edge from parent[draw=black,thin]
    }
    edge from parent[draw=red, very thick]
}
      }
    child { node {$I_8$}  } 
}
;
\end{tikzpicture}
\caption{Laminar tree representation $\lamf_\ver(t)$ of the intervals in \cref{fig:intervals}.}
\end{subfigure}
\caption{Illustration of the online intervals of a laminar online forest at time $t = 9$. In this example, we have nodes $\node_1, \ldots, \node_8$ with arrival times $a_\ver(\node_i) = i$. The algorithm has matched $(r(\node_3),r(\node_4))$ at time $4.5$, $(r(\node_5), r(\node_6))$ at $8$, where the latter leads to the termination of $\node_7$ and the creation of $\node_8$ such that $r(\node_8)=r(\node_7)$.  At $t = 9$,  requests $r(\node_1)$, $r(\node_2)$ and $r(\node_8)$ are still pending.}
\label{fig:laminar}
\end{figure}

It is easy to see that the following properties are maintained after every event.
\begin{proposition}
    \label{tree structure}
    Given an online algorithm that satisfies \cref{property: semi-laminar}, 
    for each time $t$, the set of intervals $I_\node(t)$ and $\lamf_\ver(t)$ satisfy the following properties:
    \begin{enumerate}[nosep]
        \item for every pending request $q$, there is exactly one live node $\node$ such that $r(\node)=q$;
        \item the set of live nodes of pending requests forms a chain and corresponds to a rooted subpath in $\lamf_\ver(t)$;
        \item for each $i$, the $(i+1)$-th latest pending request's live node is the parent of the $i$-th latest pending request's live node;
        \item when a request $q$ arrives at time $t$ and it is not the only pending request, a new node $\node$ such that $r(\node)=q$ becomes a child of the previous latest pending request's live node; if $q$ is the only pending request, then a new node $\node$ such that $r(\node)=q$ becomes a new root in $\lamf_\ver(t)$.
    \end{enumerate}
\end{proposition}

\cref{tree structure} immediately implies that the corresponding requests of each node and its children satisfy the following ordering, which will be used later in the analysis.
\begin{corollary}
\label{lemma:order on the chain}
    For every node $\node\in\Node_\ver$ and $\node'\in C_\node(t)$ at time $t$ when both $\node$ and $\node'$ are live, it holds that $r(\node)\prec r(\node')$.
\end{corollary}

\subsection{Description of Algorithm and Potential Function}
\label{sec:counters-matching}

In this part, we describe the sub-routine $\ALG_\ver$ of $\ALG$ for every vertex $\ver$ of $\mathcal{T}$. $\ALG_\ver$ is based on the previously defined laminar trees of interval nodes according to the requests that appear in $\mathcal{I}_\ver$. We first describe our algorithm in terms of the potential function $Y_\node$ for each node $\node$, and then define the potential function.

\paragraph{Odd and even states.} At a high level, our algorithm associates each node with a counter. At any point in time, $\ALG_\ver$ can either be in \emph{odd} or \emph{even} states, determined externally by the number of pending requests positioned at descendants of $\ver$. If this number is even, then the latest node is \emph{active} and all other nodes are \emph{frozen}; otherwise if this number is odd, then the second latest node is active and all other nodes are frozen. Note that if $\ver$ is a leaf of $\mathcal{T}$, then $\ALG_\ver$ is always in the even state and so the latest counter is active, which is the special case of a single-location metric as discussed in~\cref{sec:techniques}. The algorithm then matches either the corresponding requests of the latest two nodes or those of the second and the third latest nodes, based on the potential function. Based on those two states of $\ALG_\ver$ and the corresponding rules of increasing the counter, we have the following definition:

\begin{definition}[Critical parent and child]
    At time $t$, if $\ALG_\ver$ is in even state, then the latest two live nodes are said to be \emph{critical}; otherwise the second and third latest live nodes are critical. More specifically, among two critical nodes, the earlier one is the \emph{critical parent} and the later one is the \emph{critical child}.
\end{definition}

Since every request has a unique corresponding node at every moment when it is pending in $\mathcal{I}_\ver$, the critical nodes correspond to two pending requests at every $t$.

\paragraph{Algorithm description.}
Our algorithm associates every node $\node$ with a counter $z_\node(t)$ that evolves over time as follows. 
Denote by $z_\node(t)$ the value of $\node$'s counter at time $t$ and $z'_\node(t)$ to be its instantaneous increase.
When a node $\node$ arrives in $\mathcal{I}_\ver$, its counter is initialized to zero, i.e., $z_\node(a_\ver(\node)) = 0$. At time $t$, if $\node$ is the node that is active, then we increase its counter at the same rate as its corresponding request's delay in $\mathcal{I}_\ver$, i.e., the instantaneous counter increase is $z'_\node(t)=f'_{\ver,r(\node)}(t)$; every remaining node's  counter is frozen. Recall that we call the active node critical child and its parent critical parent (if its parent exists). We now specify the algorithm's matching condition in terms of the counter values and the potential functions $Y_\node$. Suppose we are at time $t$ and $\node$ is the critical parent. The algorithm matches $r(\node)$ with the request of the critical child when
\begin{equation}\label{condition}
    Y_\node(t)\geq \frac{1}{2}z_\node(t).
\end{equation}
And we call $\node$ a \emph{primary node} for every such match.

\begin{remark}
    We remark that the two failed attempts discussed in \cref{sec:intro} can also be defined in this framework, by setting the potential function $Y_r(t)$ to be $z_{r'}(t)$ ($r'$ is the current pending child) and $\sum_{r_{j}\in C_{r}(t)}z_{r_j}(t)$ respectively. In the single-location case, every request has a unique node at all times and so we can use the notation of requests to express nodes without ambiguity.
\end{remark}

\paragraph{Potential function $Y_\node$.}
At a high level, the potential $Y_\node(t)$ of node $\node$ at time $t$ depends on the counters of its children, and when those children were previously terminated when $\node$ was not the critical parent.

The key insight underlying the potential function is as follows. Let $\node_i\prec_\ver \node_j$ be two children of $\node$, and $\OPT_\ver$ be a laminar optimal solution (\cref{non-crossing}) of the consolidated instance induced within $T_\ver$. If
 $\OPT_\ver$ matches some request $q$ that arrives before $a_\ver(\node_i)$ with one that arrives after $m_\ver(\node_j)$, then we can charge the delay increase of $r(\node)$ between $a_\ver(\node_i)$ and $m_\ver(\node_j)$ to the match of $q$ made by $\OPT_\ver$. %
 By laminarity, we can find distinct matches of $\OPT_\ver$ that induce a partition of $\node$'s children,
which naturally suggests a charging scheme to $\OPT_\ver$; we prove this formally in \cref{sec:misalign-charging}. %
Since $\OPT_\ver$ and thus the partition is unknown to our online algorithm, we define the potential function to minimize among all. %

\begin{definition}[Canonical partitions]
Let $\node$ be a node of $\lamf_\ver(t)$ that is pending at time $t$ and $\node_1 \prec_\ver \ldots \prec_\ver \node_k$ be its current children in $C_\node(t)$.
A partition $\mathcal{P}_\node$ of $C_\node(t)$ is \emph{contiguous} if every part of the partition consists of consecutive children $\node_i, \node_{i+1}, \ldots, \node_j$, and moreover, is \emph{canonical}
if $\{\node_k\}$ is one of its parts.
\end{definition}

Given a canonical partition $\mathcal{P}_\node$ of $C_\node(t)$, its \emph{value} is defined to be the sum of the values of its parts, where the value of part  $B_\node=\{\node_{i},\ldots,\node_{j}\}$ is defined as follows:
\begin{equation}
    v_\node(B_\node)=
    \begin{cases}
        f_{\ver,r(\node)}(m_\ver(\node_{j}))-f_{\ver,r(\node)}(a_\ver(\node_{i}))&\text{if $|B_\node|>1$;} \\
        z_{\node_{j}}(m_\ver(\node_{j}))&\text{if $|B_\node|=1$,}
    \end{cases}
\end{equation}

where $m_\ver({\node_{j}})$ is assigned a tentative value of $t$ if $\node_{j}$ is still alive at time $t$. In other words, if $B_\node$ is a singleton, then its value is simply the counter value of its unique child. Otherwise, its value is the delay increase of $r(\node)$ between the moments when $\node_i$ is added to the tree and when the last child $\node_j$ in $B_\node$ is terminated.

\begin{definition}
    [Potential function $Y_\node$]
    The \emph{potential function} $Y_\node$ for node $\node$ is: %
 \begin{equation}\label{definevalue}
     Y_\node(t):=\min_{\mathcal{P}_\node\text{ is a canonical partition of }C_\node(t)}\sum_{B_\node\in\mathcal{P}_\node}v_\node(B_\node).
 \end{equation}
\end{definition}

\noindent Note that at any point in time, only the counter of the critical child can increase, and so only the potential of the critical parent can increase.

\begin{remark}\label{alg remark}
    We remark that the algorithm indeed matches all pending requests, provided that $|\allreqs|$ is even and $f_r(t)$ is unbounded for every $r\in\allreqs$. Otherwise, suppose for contradiction that some requests remain unmatched forever, then they are all going to be promoted to the root of $\mathcal{T}$. Consider the latest two live nodes $\node\prec_\ver\node'$ in the instance of root $\ver$, then $\node$ is the critical parent and $\node'$ is the critical child, because there is zero request that is positioned at any other vertex. By definition, $\node'$ is the active node, so $z_{\node'}(t)$ tends to infinity, whereas $z_\node(t)$ remains bounded because $\node$ is frozen. Consequently, the matching condition must eventually be satisfied since $Y_\node(t)\ge z_{\node'}(t)$, contradicting the assumption that $\node$ and $\node'$ stay alive at the end.
\end{remark}

\section{Algorithm Analysis}
\label{sec:analysis}

We now analyze the competitive ratio of our algorithm and prove our main result \cref{constant-competitive}.

\paragraph{Section overview.} In \cref{sec:alg-prop}, we prove a key property of $\ALG_\ver$ that the counters are geometrically decreasing along chains (\cref{align:orders}). In \cref{sec:alg-to-counters}, we define a set of primary nodes $\mathcal{D}_\ver$ for each $\Node_\ver$ and show that the cost of $\ALG$ is at most $O(1)$ times the total value of the counters of all $\mathcal{D}_\ver$'s (\cref{algorithmcost}). In \cref{sec:pruning}, we further prune the set of counters from $\mathcal{D}_\ver$ into $\pr_\ver$ while only losing a constant factor of counter values (\cref{pruned primary set}). Finally, in \cref{sec:counters-to-opt}, we show that the total value of the counters of $\pr_\ver$ is at most $O(1)$ times $\OPT_\ver$, the optimal cost of the consolidated instance induced within $T_\ver$ (\cref{aligned total,m total cost}), and thus the desired competitive ratio of $\ALG$. 

\paragraph{Notation.} For brevity, for every $\node\in\Node_\ver$, we write $z_\node$, $C_\node$, and $Y_\node$ for the values of $z_\node(t)$, $C_\node(t)$, and $Y_\node(t)$ at time $t=m_\ver(\node)$, respectively (if $\node$ stays alive at the end, then define $m_\ver(\node)=\infty$). Since the counter $z_\node(t)$ is frozen once $\node$ is terminated, we have $z_\node=z_\node(m_\ver(\node))=z_\node(t)$ for every $t\ge m_\ver(\node)$. Likewise, $C_\node=C_\node(m_\ver(\node))=C_\node(t)$ and $Y_\node=Y_\node(m_\ver(\node))=Y_\node(t)$ for every $t\ge m_\ver(\node)$. %
We write $\lamf_\ver$ to mean $\lamf_\ver(t)$ where $t$ tends to infinity.

\subsection{Geometric Chain Property}
\label{sec:alg-prop}

We show that for every node $\node$ in the forest $\lamf_\ver(t)$, its counter value is always at least twice its potential (\cref{zr and cr}) and then use it to prove that the counter values along chains are geometrically decreasing (\cref{align:orders}).

\begin{lemma}\label{zr and cr}
    For every vertex $\ver\in V$ and node $\node\in\Node_\ver$, $z_\node(t)\geq 2 Y_\node(t)$ at every time $t\geq a_\ver(\node)$, and $z_\node=2 Y_\node$ if and only if $\node$ is the critical parent when it is terminated by $\ALG_\ver$. 
\end{lemma}

\begin{proof}
Fix a node $\node\in\Node_\ver$ and define the function $\Yhat_\node(t) = \max_{t' \leq t} \Yhat_\node(t')$. Since $\ALG_\ver$ terminates $\node$ at time $t$ as a critical parent if $Y_\node(t) \geq \frac{1}{2}z_\node(t)$, to prove the lemma, it suffices to show that $\Yhat_\node(t)$ is continuous for every $t \geq a_\ver(\node)$. 

Suppose $\node$ has $|C_\node| = k$ children $\node_1 \prec_\ver \ldots \prec_\ver \node_k$ and $\mP^*_\node(t)$ is the canonical partition of minimum value. For each child $\node_i$, since $C_\node(t)$ does not vary between its arrival and the arrival of the next child $\node_{i+1}$, we have that $\mP^*_\node(t)$ also does not vary until $\node_{i+1}$ arrives. Let $\mP^*_\node(i)$ be the optimal partition of $C_\node(t)$ for $t \in [a_\ver(\node_i), a_\ver(\node_{i+1}))$. Since the counter value $z_q$ is continuous for every node $q$, we also have that $Y_\node(t)$ and $\Yhat_\node(t)$ are continuous in the interval $[a_\ver(\node_i), a_\ver(\node_{i+1}))$.  

At $t = a_\ver(\node_{i+1})$, the set of canonical partitions changes, which may cause a discontinuity in $Y_\node$. Since the counter of $\node_{i+1}$ is 0 on arrival, the partition $\mP^*_\node(i) \cup \{\{\node_{i+1}\}\}$ is a canonical partition of $C_\node(a_\ver(\node_{i+1}))$ with value equal to that of $\mP^*_\node(i)$. Thus, if it is discontinuous at $a_\ver(\node_{i+1})$, the potential can only decrease and so $\Yhat_\node(t)$ remains continuous.

We conclude that $\Yhat_\node(t)$ is continuous for every $t \geq a_\ver(\node)$, as desired.
\end{proof}

We now use \cref{zr and cr} to prove that the counter values along chains are geometrically decreasing.

\begin{lemma}[Geometric Chain Property]\label{align:orders}
Let $\node_1 \prec_\ver \ldots \prec_\ver \node_k$ be all of the live nodes in $\lamf_\ver(t)$. Then $z_{\node_{j}}(t) \geq 2z_{\node_{j+1}}(t)$ and $z_{\node_{j}} \geq 2z_{\node_{j+1}}$ for every $1\leq j\leq k-1$.    
\end{lemma}

\begin{proof}
\cref{tree structure} implies that $\node_{j+1}$ is the latest child of $\node_j$ among $C_{\node_j}(t)$.
By \cref{zr and cr} and the fact that $\{\node_{j+1}\}$ is a part in every canonical partition of $C_{\node_j}(t)$,  we have that $z_{\node_j}(t) \geq 2Y_{\node_j}(t) \geq 2z_{\node_{j+1}}(t)$. This proves the first part of the lemma. 
 
For the second part of the lemma, we first note that $m_\ver(\node_j)\geq m_\ver(\node_{j+1})$ according to laminarity of node intervals. As a result, we apply the first part of the lemma to $t=m_\ver(\node_{j+1})$ and get 
$$z_{\node_{j}}(m_\ver(\node_{j+1})) \geq 2 z_{\node_{j+1}}(m_\ver(\node_{j+1}))=2z_{\node_{j+1}}. $$ 
Since counter values are nondecreasing with time, we get that $z_{\node_j} \geq z_{\node_{j}}(m_\ver(\node_{j+1}))\geq  2 z_{\node_{j+1}}$.
\end{proof}

This property of decreasing geometric sequence will play an important role in allowing us to consider a subset of nodes and charge their counters to $\OPT_\ver$, so as to avoid overcharging (\cref{lem:prune}).

\subsection{Charging $\ALG_\ver$ to Counters}
\label{sec:alg-to-counters}
In this part, we show that
we can use the total counter value of all \emph{primary requests} as a proxy for the total cost of all $\ALG_\ver$'s.%

\begin{definition}[Active and reluctant time]
For every $\node\in\Node_\ver$, %
define its \emph{active time} $J_{\node}$ to be the set
of times $t$ in which $\node$ is active, its \emph{reluctant time} $J_{\node,1}$ to be the set of times $t$ in which $\node$ is the latest live node in $\lamf_\ver(t)$ yet inactive ($\ALG_\ver$ is in odd state). We also use $J_{\node,0}$ to denote the time when $\node$ is neither active nor reluctant, or equivalently, $J_{\node,0}=I_{\node}\setminus(J_{\node}\cup J_{\node,1})$. %
\end{definition}

For each node $\node$, we are going to relate the actual delay increase during interval $I_{\node}$ of request $r(\node)$ to the final counter values of some nodes among different positions in $\mathcal{T}$. Since every pending request in $\mathcal{I}_\ver$ has a unique corresponding node interval at each moment, the total delays of all $\ALG_\ver$'s can be bounded by the sum of delay increase during every node interval of its corresponding request, which allows us to analyse in the node-wise instead of request-wise way. Furthermore, for each node, we charge its reluctant time and the remaining part separately.

We first establish the following structural property, which shows that the interval $J_{\node,0}$ can be expressed as the union of $J_q$ and $J_{q,0}$ of every of its children $q$.

 \begin{claim}\label{C-partition}
 Let $\node$ be a node. Then, for any two children $q, q'$ of $\node$, we have $I_q \cap I_{q'} = \varnothing$. Moreover, $J_{\node,0}=\cup_{q\in C_{\node}}(J_{q}\cup J_{q,0})$.
\end{claim}
\begin{proof}
Since $q$ and $q'$ are siblings, \cref{tree structure} implies that they cannot both be live at the same time, which in turn implies that $I_q \cap I_{q'} = \varnothing$.
According to our definition of two states of $\ALG_\ver$ and reluctant time of nodes, $J_{\node,0}\subseteq \cup_{q\in C_{\node}}I_q$. Moreover, at each $t\in I_q=J_q\cup J_{q,0}\cup J_{q,1}$ for a child $q\in C_\node$: if $t\in J_{q,1}$, then counter $z_\node(t)$ is active and $t\in J_{\node}$; if $t\in J_q\cup J_{q,0}$, then $\node$ is neither active nor reluctant, and so $t\in J_{\node,0}$. As a result, $J_{\node,0}=\cup_{q\in C_{\node}}(J_{q}\cup J_{q,0})$.
\end{proof}
 
\begin{lemma}\label{individual upper}
For every vertex $\ver\in V$ and $\node\in\Node_\ver$, we have $z_\node \leq f_{\ver,r(\node)}(J_{\node}\cup J_{\node,0})\leq 2z_{\node}$.
\end{lemma}
\begin{proof}
Observe that $z_\node = f_{\ver,r(\node)}(J_{\node})$, and so $z_\node\leq f_{\ver,r(\node)}(J_{\node}\cup J_{\node,0})$ is immediate. We prove $f_{\ver,r(\node)}(J_{\node}\cup J_{\node,0})\leq 2z_{\node}$ by induction over $\lamf_\ver$. %
    
We start with the base case: when $I_\node$ is a leaf in $\lamf_\ver$. In this case, $C_\node = \varnothing$, and so by \cref{C-partition}, $J_{\node,0}=\varnothing$. Thus, $z_\node = f_{\ver,r(\node)}(J_{\node}\cup J_{\node,0})$.

Next, suppose $C_\node\neq\varnothing$ and the claim holds for every child $\node'\in C_\node$. By \cref{C-partition}, we have $f_{\ver,r(\node)}(J_{\node}\cup J_{\node,0})=z_{\node}+\sum_{\node_j\in C_\node}f_{\ver,r(\node)}(J_{\node_j}\cup J_{\node_j,0})$. It remains to relate the sum to $Y_\node$, which is at most $z_\node/2$ by \cref{zr and cr}. 

Suppose $C_\node$ consists of $\node_1\prec_\ver\ldots\prec_\ver \node_k$ and $\mathcal{P}_\node^{*}$ is the canonical partition of $C_\node$ such that $\sum_{B_\node\in \mathcal{P}^*_\node}v_\node(B_\node)=Y_\node$, then we have 
$\sum_{\node_j\in C_\node}f_{\ver,r(\node)}(J_{\node_j}\cup J_{\node_j,0})
    = \sum_{B_\node\in\mathcal{P}^{*}_{\node}} \sum_{\node_j \in B_\node} f_{\ver,r(\node)}(J_{\node_j}\cup J_{\node_j,0})$.  
We now show, for each part $B_\node \in \mP^*_\node$, that  $\sum_{\node_j \in B_\node} f_{\ver,r(\node)}(J_{\node_j}\cup J_{\node_j,0})\leq 2v_\node(B_\node)$. Suppose $|B_\node|>1$, and let $\node_{b_1}$ be the earliest node and $\node_{b_2}$ the latest in $B_\node$. 
Then,
\begin{equation*}
    \sum_{\node_j \in B_\node} f_{\ver,r(\node)}(J_{\node_j}\cup J_{\node_j,0})
    \leq f_{\ver,r(\node)}(m_u(\node_{b_2}))-f_{\ver,r(\node)}(a_u(\node_{b_1}))= v_\node(B_\node).
\end{equation*}
On the other hand, if $B_\node=\{\node_b\}$, then because $r(\node)\prec r(\node_b)$ (\cref{lemma:order on the chain}) and delays are noncrossing, we have
\begin{equation*}
    f_{\ver,r(\node)}(J_{\node_b}\cup J_{\node_b,0})\leq f_{\ver,r(\node_b)}(J_{\node_b}\cup J_{\node_b,0})\leq 2z_{\node_b}=2v_\node(B_\node),
\end{equation*}
where the last inequality is obtained by applying the inductive hypothesis on $\node_{b}$. Thus, we have shown that $\sum_{\node_j \in B_\node} f_{\ver,r(\node)}(J_{\node_j}\cup J_{\node_j,0})\leq 2v_\node(B_\node)$ for each part $B_\node \in \mP^*_\node$.

Putting the above together, we have
\begin{align*}
    f_{\ver,r(\node)}(J_{\node}\cup J_{\node,0})&=z_{\node}+\sum_{B_\node\in\mathcal{P}^{*}_{\node}} \sum_{\node_j \in B_\node} f_{\ver,r(\node)}(J_{\node_j}\cup J_{\node_j,0})\\
    &\leq z_{\node}+2\sum_{B_\node\in\mathcal{P}^{*}_{\node}} v_\node(B_\node)\\
    &= z_{\node}+2Y_\node\leq 2z_{\node},
\end{align*}
where the last inequality holds by \cref{zr and cr}.
\end{proof}

Next we consider the delay increase during $J_{\node,1}$ when $\node$ is reluctant. In this part, we want to avoid nodes $\node$ such that: $\node\in\lamf_\ver(t)$ but $r(\node)$ is not positioned at $\ver$, because they do not actually represent their requests' positions or contribute to positive counter increase. At time $t$, we call a live node $\node\in \Node_\ver$ a \emph{valid node} if $r(\node)$ is a pending request positioned at $\ver$. We then show that when a valid node $\node$ is reluctant at time $t$, we can always find another valid node that is not reluctant at some moment within the subtree $T_\ver$ to charge the instantaneous delay increase of $r(\node)$ to. Moreover, the total charge a valid node that is not reluctant at a given moment can receive is no more than $H$ times its instantaneous delay increase. In summary, the following lemma gives us an upper bound of the total delay incurred by every node when it is reluctant, expressed in terms of the total counter values.
\begin{lemma}
\label{lemma: subtree charge}
$\sum_{u\in V}\sum_{\node\in\Node_u}f_{\ver,r(\node)}(J_{\node,1})\leq 4(H-1)\sum_{u\in V}\sum_{\node\in\Node_u}z_\node$.
\end{lemma}
\begin{proof}
 We will define a pair of mappings $(\zeta,\chi)$ on all tuples $(\node,t)$ of valid node $\node$ and time $t$ such that $\node$ is reluctant at $t$. Furthermore, $\zeta(\node,t)$ is a tuple $(\node',t')$ such that $\node'$ is a valid node at $t'$, and $\chi(\node,t)$ is a value such that $f_{r(\node)}'(t)\leq \chi(\node,t)f'_{r(\node')}(t')$.

Now we show how to define the mapping. Given a valid reluctant $\node$ at time $t$ whose position is $\ver$, the total number of requests positioned at descendants of $\ver$ is odd. As a result, there is a descendant $u'$ of $u$ such that: at time $t$, there is at least one request positioned at $u'$ and the total number of requests positioned with $T_{u'}$ is odd, while no request is positioned at any vertex on the path between $u$ and $u'$. If there is at least one valid active node at $u'$, then define $\zeta(\node,t)=(\node',t)$ and $\chi(\node,t)=1$. Otherwise, we show that there must be a request $q$ positioned at $\ver'$ that does not have a live node: (Case 1) if an odd number of requests are positioned at $\ver'$, then $\ALG_{\ver'}$ is in even state and so as long as there exists at least one valid node, the latest one must be active; (Case 2) if an even number (at least two) of requests are positioned at $\ver'$, then $\ALG_{\ver'}$ is in odd state and so the second latest live (and valid) node is active if every request positioned at $\ver'$ has a live node. As a result, $\ALG_{u'}$ must have matched $q$ with a request $q'\prec q$ such that $q'$ has already been promoted to the parent of $u'$ before this match. In this case, let $t'$ be the moment that $f_{q'}(t')=f_{q}(t)$ ($t'$ is unique according to our bijection assumption of every request's delay), then $q'$ has a unique valid node $\node'$ in $\lamf_{u'}(t')$. We define $\zeta(\node,t)=(\node',t')$ and $\chi(\node,t)=\frac{f'_{q}(t)}{f'_{q'}(t')}$. It is easy to see that $f'_{r(\node)}(t)\leq \chi(\node,t)f'_{r(\node')}(t')$ holds as $f_{r(\node)}'(t)\leq f'_{q}(t)$ (\cref{obs: promotion}), and we have the local derivative $\frac{dt'}{dt}=\chi(\node,t)$.

From the above paragraph, $\zeta(\node,t)$ is well-defined for every reluctant valid node at $t$. Moreover, if $(\node',t')=\zeta(\node,t)$, then the position of $\node'$ is a descendant of that of $\node$. Because there is no reluctant node at any leaf vertex, for every reluctant $\node$ at time $t$, there exist $k\leq H-1$ and $\zeta^k(\node,t)=(\node',t')$ such that $\node'$ is not a reluctant node at time $t'$. We define $\mathcal{G}(\node,t)=(\node',t')$ and $\Psi(\node,t)=\Pi_{\ell=0}^{k-1} \chi(\zeta^\ell(\node,t))$, then $f_{r(\node)}'(t)\leq \Psi(\node,t)f'_{r(\node')}(t')$ and so we charge the delay increase of $f'_{r(\node)}(t)dt$ to $f'_{r(\node')}(t')dt'$ by the substitution $dt'=\Psi(\node,t)dt$ derived from the chain rule of differentiation.

Next we bound the number of tuples $(\node,t)$ such that $\mathcal{G}(\node,t)=(\node',t')$ for every valid non-reluctant node $\node'$ at time $t'$. $|\zeta^{-1}(\node',t')|\leq 2$ according to the two cases that we define $\zeta(\node,t)$. For every reluctant node $\node$ at time $t$, $|\zeta^{-1}(\node,t)|$ is at most one, because in the first case the function value $(\node,t)$ has to be an active node $\node$ at time $t$. As a result, there are at most two different chains beginning at $(\node',t')$ where the next element in the chain is a pre-image of the previous one, and each chain is of length no more than $H$ (including $(\node',t')$ itself). Therefore, $|\mathcal{G}^{-1}(\node',t')|=\sum_{k=1}^{H-1}|\zeta^{-k}(\node',t')|\leq 2(H-1)$.

We conclude that the total charge is no more than $2(H-1)\sum_{u\in V}\sum_{\node\in\Node_u}f_{\ver,r(\node)}(J_{\node}\cup J_{\node,0})$, which is no more than $4(H-1)\sum_{u\in V}\sum_{\node\in\Node_u}z_\node$ according to~\cref{individual upper}.
\end{proof}

Combined with \cref{individual upper}, it holds that $\sum_{\ver\in V}\ALG_\ver \leq (4H-2)\sum_{\ver\in V}\sum_{\node \in \Node_\ver} z_\node$.

Next, we show that it suffices to consider the counters of primary nodes. Let $\md_\ver$ denote the set of primary nodes in $\Node_\ver$: we have previously defined every critical parent node when a match happens to be a primary node, here we complement it by also including the earliest node among all that stay alive at the end (if there is any).
By applying \cref{align:orders} to nodes that are terminated at the same moment and nodes that stay alive at the end, it is easy to see that we can use $\md_\ver$ to represent node counters in $\Node_\ver$ while only losing a factor of $2$. Thus, we can bound the cost of all single-location subroutines by the counters of primary nodes, up to a constant:

\begin{lemma}
[Charging to Counters]\label{algorithmcost}
    $\sum_{\ver\in V}\ALG_\ver\leq (8H-4)\sum_{\ver\in V}\sum_{\node \in \md_\ver} z_\node$.
\end{lemma}

\subsection{Pruning Counters}
\label{sec:pruning}

We now fix a vertex $\ver$ of $\mathcal{T}$ and prune the set of counters from $\mathcal{D}_\ver$ based on $\OPT_\ver$'s matching $M^*_\ver$, to finally define the desired set $\pr_\ver$ that we will charge to $\OPT_\ver$. 

For the case where there is one request $r$ unmatched by $\OPT_\ver$ because $|\mathcal{R}_\ver|$ is odd, we define a request $r_\infty$ such that $a(r_\infty)=\infty$ and its delay is zero, and include $(r,r_\infty)\in M^*_\ver$. For every time interval $I$, we say that $r_\infty\in I$ if $I$ is right-unbounded. In the definitions and analysis hereafter (e.g. \cref{preempted,def: match rho}) we regard $r_\infty$ as a request in $\mathcal{R}_\ver$ and $(r,r_\infty)$ as a match made by $\OPT_\ver$.

In the following analysis and definitions, for an interval $I_\node$, if $\node$ is at a leaf vertex $u$ and so $a(r(\node))=a_\ver(\node)$, then we do not include the request $r(\node)$ itself when we consider $R_\ver(I_\node)$. Otherwise, we assume without loss of generality that no request arrival occurs at the exact moment $a_\ver(\node)$.

\begin{definition}
    [Preempted nodes and intervals]
    \label{preempted}
    A node $\node\in\Node_\ver$ and its interval $I_\node$ are said to be \emph{left-preempted} (or \emph{right-preempted}, resp.) if there exists a request $r\in \mathcal{R}_\ver$ that arrives during $I_\node$ that $\OPT_\ver$ matches with a request $q$ with $a(q) < a_\ver(\node)$ (or $a(q) > m_\ver(\node)$, resp.).
\end{definition}
\noindent Note that a node may be simultaneously left- and right-preempted.

\begin{definition}
    [Left- and right-equivalent requests]
    For every $\node\in\Node_\ver$, if $\node$ is left-preempted, define $\hl(\node)$ to be the latest arrival in $R_\ver(I_\node)$ that is matched by $\OPT_\ver$ to a request that arrives before $a_\ver(\node)$. Similarly, if it is right-preempted, define $\hr(\node)$ to be the earliest in $R_\ver(I_\node)$ that is matched by $\OPT_\ver$ to a request that arrives after $m_\ver(\node)$. Nodes $\node_1,\node_2$ are said to be \emph{left-equivalent} if $\hl(\node_1) = \hl(\node_2)$, and \emph{right-equivalent} if $\hr(\node_1)=\hr(\node_2)$.
\end{definition}

Later, we will charge portions of the counter values to $\OPT_\ver$ based on $\hl(\node)$ and $\hr(\node)$. This may lead to overcharging if multiple requests belong to the same left- or right-equivalence class. The following lemma allows us to restrict our attention to a pruned subset in which no two requests are left- or right-equivalent, while losing only a constant factor.

\begin{lemma}
\label{lem:prune}
For any set of nodes $\mathcal{S}\subseteq \Node_\ver$, there exists a subset $\mathcal{S}^{\mathrm{p}} \subseteq \mathcal{S}$ such that 
$\sum_{\node \in \mathcal{S}} z_\node\leq  4\sum_{\node\in \mathcal{S}^{\mathrm{p}}} z_\node$ and no two nodes in $\mathcal{S}^{\mathrm{p}}$ are left-equivalent or right-equivalent.
\end{lemma}

\begin{proof}
First, observe that if two nodes $\node_1,\node_2$ are left-equivalent or right-equivalent, then either $I_{\node_1} \supseteq I_{\node_2}$ or $I_{\node_1} \subseteq I_{\node_2}$. In particular, a left-equivalence (or right-equivalence) class $Q$ forms a chain, and thus \cref{align:orders} implies that $\sum_{\node \in Q} z_\node\leq 2 \max_{\node \in Q} z_\node$.

Given any $\mathcal{S}$, we obtain $\mathcal{S}^{\mathrm{p}}$ in two stages. Initially, let $\mathcal{S}^{\mathrm{p}}=\mathcal{S}$. In the first stage, for each left-equivalence class in $\mathcal{S}^{\mathrm{p}}$, we remove from $\mathcal{S}^{\mathrm{p}}$ all nodes except the one with the highest counter value. In the second stage, we apply the same pruning procedure to each right-equivalence class among the remaining nodes in $\mathcal{S}^{\mathrm{p}}$.

By the above argument, the total counter value of $\mathcal{S}^{\mathrm{p}}$ decreases by a factor of at most $2$ in each stage. Hence, $\sum_{\node \in \mathcal{S}} z_\node\leq  4\sum_{\node\in \mathcal{S}^{\mathrm{p}}} z_\node$.
\end{proof}

Combining \cref{lem:prune,algorithmcost}, we immediately get the following:
\begin{proposition}
\label{pruned primary set}
    There is a set of nodes $\pr_{\ver}\subseteq \md_\ver$ such that the total cost of all virtual algorithms $\sum_{\ver\in V}\ALG_\ver\leq (32H-16)\sum_{\ver\in V}\sum_{\node\in\pr_\ver}z_\node$ and no two nodes in $\md^{\mathrm{p}}_\ver$ are left-equivalent or right-equivalent.
\end{proposition}
\subsection{Charging Counters to $\OPT_\ver$}
\label{sec:counters-to-opt}

Fix $\ver\in V$. Let $\pr_\ver$ be the pruned set of primary nodes guaranteed by \cref{pruned primary set}.
In the remainder of this section, we show that $\sum_{\node\in\pr_\ver} z_{\node} \leq O(1)\cdot\OPT_\ver$ via a charging scheme. We first define odd and even subintervals which lets us partition nodes into two groups: aligned and misaligned, based on the state of $\ALG_\ver$ right after their arrivals as well as how $\OPT_\ver$ matches requests that arrive during those intervals. In \cref{align charging}, we show that the total counter value of aligned nodes is at most $O(1)\cdot\OPT_\ver$. Handling the misaligned nodes is significantly more challenging, and is where the potential function becomes necessary. In \cref{sec:misalign-charging}, we charge part of the counters to aligned nodes and some directly to $\OPT_\ver$.

\begin{definition}[Odd and even nodes]
\label{def: mode on arrival}
For every $\node\in\Node_\ver$, consider the state of $\ALG_\ver$ right after $a_\ver(\node)$ when $\node$ arrives: we say that $\node$ is odd (or even, resp.) if $\ALG_\ver$ is in odd (or even, resp.) state.
\end{definition}
\noindent Note that if $\node$ is at a leaf vertex, then it is always even, which is a special case where the algorithm runs on a single-location metric.

\paragraph{Odd and even subintervals.} 
We use $I_{\node}^{\mathrm{odd}}$ to denote all time $t\in I_{\node}$ by which an odd number of requests from $R_\ver(I_{\node})$ have arrived and similarly define $I_{\node}^{\mathrm{even}}$. We get that whenever the counter of node $\node$ is active at $t$, it must hold that $t\in I^{\mathrm{odd}}_\node$ if $\node$ is odd and $t\in I^{\mathrm{even}}_\node$ if $\node$ is even (\cref{odd counter}). As a result, when we charge $z_\node$ to $\OPT_\ver$, we want to find a request that is pending during $I^{\mathrm{odd}}_\node$ or $I^{\mathrm{even}}_\node$ depending on the state of $\ALG_\ver$ right after $a_\ver(\node)$, and we will discuss different cases depending on whether such a request can be found.

\begin{definition}
[$I_{\node}^{\mathrm{odd}}, I_{\node}^{\mathrm{even}}$]
\label{parity division}
Let $\node$ be a node in $\Node_\ver$ and $k = |R_\ver(I_\node)|$, the number of requests that arrive during time interval $I_\node$ within subtree $T_\ver$. For each $1 \leq j \leq k$, let $r_j$ be the $j$-th earliest request of $R_\ver(I_\node)$ and $b_j$ be its arrival time (define $b_0$ to be $a_\ver(\node)$).
Moreover, define a request $r_j$ to be \emph{$\node$-odd} if $j$ is odd and \emph{$\node$-even} otherwise.
The arrival times $a_\ver(\node)=b_0<b_1 < \ldots < b_{k} < m_\ver(\node)$ partition $I_\node$ into odd and even subintervals as follows: for $j< k$, the $j$-th subinterval is $[b_j, b_{j+1})$; the final subinterval is $[b_{k}, m_\ver(\node))$. Denote by $I_{\node}^{\mathrm{odd}}$ and $I_{\node}^{\mathrm{even}}$ the union of the $j$-th subintervals for odd $j$ and even $j$, respectively.
\end{definition}

In the remainder of the analysis, when we compare $\ALG_\ver$ and its node counters and intervals against all requests in $\mathcal{R}_\ver$, we regard those that have not been promoted to $\ver$ as requests under $\ALG_\ver$ as well. For requests that have not been promoted to $\ver$, we regard their matching status under $\ALG_\ver$ as per whether they are pending or have been matched by the \textbf{actual} algorithm $\ALG$. After they are promoted to $\ver$, we regard their matching status under $\ALG_\ver$ as per whether they have been matched in the \textbf{virtual} instance $\mathcal{I}_\ver$, rather than whether they have actually been matched by $\ALG$. According to the mechanism of each $\ALG_\ver$ and \cref{lemma:order on the chain}, the counter of a live node $\node$ is only active when the number of pending requests under $\ALG_\ver$ that arrive later than $r(\node)$ is even. The next lemma derived from basic properties of $\ALG_\ver$ and the laminar forest relates its parity to the arrivals of requests in $\mathcal{R}_\ver$.

\begin{lemma}
\label{lemma: interval match parity}
For every $\node\in\Node_\ver$, during $I_\node$, the parity of the number of pending requests under $\ALG_\ver$ that arrive after $r(\node)$ only changes when a new request in $\mathcal{R}_\ver$ arrives.
\end{lemma}
\begin{proof}
It is easy to see the parity in question does not change when a request is promoted to a new vertex within $T_\ver$, or when two requests that are positioned at some descendant of $\ver$ are matched by $\ALG$. If it is changed by a match made by $\ALG_\ver$ in $\mathcal{I}_\ver$, then $\ALG_\ver$ must have matched a pair of requests $(q,q')$ that $q\prec r(\node)\prec q'$. Assume this happens at time $t$, then $\node$ is between nodes of $q$ and $q'$ on the path of live nodes in $\lamf_\ver(t)$ (\cref{tree structure}), a contradiction to \cref{property: semi-laminar}.
\end{proof}
Now we come to the key property that links the counter increase to the parity of subintervals.
\begin{lemma}
\label{odd counter}
For every node $\node\in\Node_\ver$, we have
    $J_{\node}\subseteq I_{\node}^{\mathrm{odd}}$ if $\node$ is odd and $J_{\node}\subseteq I_{\node}^{\mathrm{even}}$ if $\node$ is even.
\end{lemma}
\begin{proof}
We prove the case that $\node$ is odd, and the other case can be shown analogously. At each moment $t\in J_\node$, there are even number of pending requests in $\mathcal{R}_\ver$ under $\ALG_\ver$ that arrive after $r(\node)$. Because $\node$ is odd, the parity in question is odd at $a_\ver(\node)$. Furthermore, it only changes when a new request in $\mathcal{R}_\ver$ arrives (\cref{lemma: interval match parity}), and so for every moment $t$ when it is even, $t$ must be in $I^{\mathrm{odd}}_\node$, hence our claim.
\end{proof}

We now consider how the requests in $R_\ver(I_{\node})$ are matched by $\OPT_\ver$, which will be important for our charging scheme.

\begin{definition}
    [Aligned and misaligned nodes]
    A node $\node\in\Node_\ver$ is \emph{aligned} if: $\hl(\node)$ is $\node$-odd (or $\node$-even, resp.) and $\node$ is even (or odd, resp.); 
    or, if $\node$ is odd and not left-preempted. Otherwise $\node$ is \emph{misaligned}. 
Let $\pr_{\ver,a}$ and $\pr_{\ver,m}$ be the sets of aligned and misaligned nodes in $\pr_\ver$, respectively.
\end{definition}

We need to be cautious with the primary node (at most one) that stays alive at the end, because it is not matched as a critical parent and hence we cannot use its potential to charge, as what we do to misaligned primary nodes in \cref{sec:misalign-charging}. Luckily, we show that such a primary node must be aligned and so we can directly charge $z_\node$ to $\OPT_\ver$ as in \cref{align charging}, without reliance on $Y_\node$. 

We first prove a property that will be used for $\OPT_\ver$, as it is laminar.
\begin{lemma}\label{even}
    Let $(M,m)$ be a laminar solution. Then, for every match $(r,r') \in M$, the number of requests $q$ such that $r \prec q \prec r'$ is even.
\end{lemma}
\begin{proof}
Because a laminar solution always matches the latest two pending requests at the time, when $(r,r')$ is matched, all requests that arrive between $r$ and $r'$ have all been matched. Any request $q$ that $r\prec q\prec r'$ cannot be matched with any request $q'$ such that $q'\prec r$ or $q'\succ r'$ when $r$ and $r'$ are both unmatched, otherwise it is a violation of the laminar property. As a result, all requests that arrive between $r$ and $r'$ are matched into pairs by $(M,m)$, and thus the number of such requests must be even.
\end{proof}

\begin{lemma}
\label{lemma: misaligned primary}
If $\node\in\Node_\ver$ is the primary node that stays alive at the end, then $\node$ is aligned.
\end{lemma}
\begin{proof}
Because $\node\in\Node_\ver$ is the primary node that stays alive at the end, all requests that arrive before $r(\node)$ are: either never promoted to $\ver$, or matched in $\mathcal{I}_\ver$ before $a_\ver(\node)$. Let $k_1$ be the number of requests that arrive before $r(\node)$, $k_2$ be the number of requests that arrive during $(a(r(\node)),a_\ver(\node)]$, $k_3$ be the number of requests that arrive during $(a(r(\node)),a_\ver(\node)]$ and are unmatched under $\ALG_\ver$ at $a_\ver(\node)$. According to the mechanism of $\ALG_\ver$ and \cref{lemma: interval match parity}, for requests that arrive before $r(\node)$: under $\ALG_\ver$, either they are matched with each other, or they can only be matched with requests that arrive during $(a(r(\node)),a_\ver(\node)]$ by the moment $a_\ver(\node)$. Therefore, $k_1-k_2$ has the same parity as $k_3$, and $\node$ is odd (or even, resp.) if and only if $k_3$ is odd (or even, resp.); this happens if and only if $k_1+k_2+1$, the number of requests that arrive before $a_\ver(\node)$ (including $r(\node)$), is even (or odd, resp.). By laminarity of $\OPT_\ver$ and \cref{even}, when $\hl(\node)$ exists, because $\rho(\hl(\node))$ is at an odd position among all requests in $\mathcal{R}_\ver$, $\hl(\node)$ is at an even position. If the number of requests that arrive before $a_\ver(\node)$ is even (or odd, resp.), then $\hl(\node)$ is $\node$-even (or $\node$-odd, resp.). If $\hl(\node)$ does not exist, then $k_1+k_2+1$ must be even and so $\node$ must be odd. In summary, $\node$ must be aligned.
\end{proof}

We introduce the following notation to express the structure of $\OPT_\ver$ as well as its cost.

\begin{definition}
\label{def: match rho}
    For every $r\in\allreqs_\ver$ that is matched by $\OPT_\ver$, let $\rho(r)$ denote the request that $\OPT_\ver$ matches $r$ with, and $c(r)$ denote the delay incurred by a request $r$ under $\OPT_\ver$ if $\rho(r)\succ r$ or the delay incurred by $\rho(r)$ if $r\succ \rho(r)$, both under $\OPT_\ver$.
\end{definition}
\noindent Observe that $\OPT_\ver=\sum_{r\in\allreqs_\ver:r\prec \rho(r)}c(r)=\sum_{r\in\allreqs_\ver:r\succ \rho(r)}c(r)$.

\paragraph{Dividing $I_\node$ and $J_\node$ into two parts.} If $\node$ is left-preempted, let $\ql(\node)$ be the first request that arrives after $\hl(\node)$; otherwise, define $\ql(\node)$ to be the first request that arrives after $a_\ver(\node)$. We divide $I_\node = [a_\ver(\node), m_\ver(\node))$ into two parts: $\Il_\node = [a_\ver(\node), a(\ql(\node)))$ and $\Ir_\node = [a(\ql(\node)), m_\ver(\node))$ ($\Ir_\node=\varnothing$ if $\ql(\node)$ arrives after $m_\ver(\node)$ or does not exist). Observe that $R_\ver(\Il_\node)$ is the subset of $R_\ver(I_\node)$ that arrive before $\ql(\node)$ (including $\hl(\node)$ if $\node$ is left-preempted). Moreover, if $\node$ is not left-preempted, then all requests in $R_\ver(I_\node)$ arrive during $\Ir_\node$; if $\ql(\node)\notin R_\ver(I_\node)$ or $\ql(\node)$ does not exist, then all requests in $R_\ver(I_\node)$ arrive during $\Il_\node$. We also divide $J_\node$ into $\Jl_\node = J_\node \cap \Il_\node$ and $\Jr_\node = J_\node \cap \Ir_\node$.

\begin{claim}\label{hr block}
$\OPT_\ver$ matches every request in $R_\ver(\Ir_\node)$ to one that arrives no earlier than $\ql(\node)$.
\end{claim}

\begin{proof}
By definition of left-preemption, 
the claim is true if $\node$ is not left-preempted. Suppose $\node$ is left-preempted.
Since every request of $R_\ver(\Ir_\node)$ arrives after $\hl(\node)$, by definition of $\hl(\node)$, no request in $R_\ver(\Ir_\node)$ is matched by $\OPT_\ver$ to a request that arrives earlier than $a_\ver(\node)$. Moreover, $\OPT_\ver$ matches $\hl(\node)$ to a request that arrives earlier than $a_\ver(\node)$. Thus, by laminarity, no request in $R_\ver(\Ir_\node)$ can be matched with a request in $R_\ver(\Il_\node)$ either. This gives us the claim.
\end{proof}

\subsubsection{Charging aligned counters $\pr_{\ver,a}$ to $\OPT_\ver$}\label{align charging}

Recall that node $\node$ is aligned if $\hl(\node)$ is $\node$-odd (or $\node$-even, resp.) and $\node$ is even (or odd, resp.), or if $\node$ is odd and not left-preempted. In this subsection, we are going to prove a stronger statement. Let $\Node_{\ver,a}$ be the set of all aligned nodes in $\Node_\ver$ and $\Node_{\ver,a}^{\mathrm{p}}$ be the pruned subset of it by applying \cref{lem:prune} to $\Node_{\ver,a}$, and so $\sum_{\node\in \Node_{\ver,a}}z_\node\leq 4\sum_{\node\in\Node_{\ver,a}^{\mathrm{p}}}z_\node$. We will charge $\Node_{\ver,a}^{\mathrm{p}}$ to $\OPT_\ver$, which then provides an upper bound of $\Node_{\ver,a}$ as well as $\pr_{\ver,a}$ since $\pr_{\ver,a}\subseteq \Node_{\ver,a}$. Our main lemma is the following:
\begin{lemma}
[Bounding Aligned Counters]
\label{aligned total}
$\sum_{\node\in\md_{\ver,a}^{\mathrm{p}}}z_\node\leq 8\OPT_\ver$.    
\end{lemma}
\paragraph{High level overview.}
The key property of an aligned node $\node$ that is odd (or even, resp.) is that, for every odd (or even, resp.) subinterval of $I_{\node}$ from $\ql(\node)$ onwards, $\OPT_\ver$ has at least one request in $R_\ver(\Ir_\node)$ that is pending for the entire subinterval. Since $\node$ can only be active during odd (or even,resp.) subintervals if $\node$ is odd (or even, resp.), for $t \geq a(\ql(\node))$, we can charge the instantaneous delay of $r(\node)$ when $\node$ is active to the instantaneous delay of a pending request in $\OPT_\ver$. Since we can find such a pending request in $R_\ver(I_\node)$, its instantaneous delay increase must always be at least that of $r(\node)$ by noncrossing delays, and thus the instantaneous charging argument works. We can charge the remaining portion of the counter value, i.e., the part up till $a(\ql(\node))$, to $c(\hl(\node))$, whose value is at least $f_{\ver,r(\node)}(a(\hl(\node)))-f_{\ver,r(\node)}(a_\ver(\node))$ no matter whether $\rho(\hl(\node))$ arrives before or after $r(\node)$, also by properties of noncrossing delays, and thus at least $z_\node(a(\hl(\node)))$.

\begin{claim}\label{chargable}
Let $\node$ be an aligned node such that $\ql(\node)\in R_\ver(I_\node)$. If $\node$ is odd (or even, resp.), then for every odd (or even, resp.) subinterval $I = [b_j, b_{j+1})$ of $I_\node$ with $b_j \geq a(\ql(\node))$, there exists a request $g_\node(t) \succeq r(\node)$ that is pending in $\OPT_\ver$ for every $t \in I$.
\end{claim}
\begin{proof}
Let $S$ be the set of requests that arrive in $[a(\ql(\node)), b_j]$.
Since $\node$ is aligned, regardless of whether $\node$ is left-preempted or not, $\ql(\node)$ is $\node$-odd (or $\node$-even, resp.) if $\node$ is odd (or even, resp.). Thus, $|S|$ is odd in both cases. By \cref{hr block}, $\OPT_\ver$ matches none of $S$ to requests earlier than $a(\ql(\node))$. Since the next request arrives at time $b_{j+1}$, $\OPT_\ver$ must keep some request of $S$ pending during the subinterval $I$. 
\end{proof}

\begin{figure}[h]
\centering
\begin{tikzpicture}
\draw (0,0) -- (7,0);
\foreach \x in {1.3,2.1,2.7,3.1,3.7,4.5,5.1,5.5,6.1} {
        \fill (\x,0) circle (2pt);
    }
\fill[red] (0.5,0) circle (2pt);
\node at (0.5,-0.32) {$\node$};
\fill (1.3,0) circle (2pt)node[font=\scriptsize][below] {$\hl(\node)$};
\node at (2.1,-0.32) [font=\footnotesize]{$p_1$};
()
\node at (4.5,-0.32) [font=\footnotesize]{$p_2$};
\draw[dashed] (0.5,0) .. controls (0.8,1) .. (6.5,1);
\draw (0,0.9)..controls (1.1,0.9)..(1.3,0);
\draw (2.1,0) parabola bend (2.9,0.8) (3.7,0);
\draw (4.5,0) parabola bend (5.3,0.8) (6.1,0);
\end{tikzpicture}
\caption{Illustration of a node $\node\in\mathcal{\Node}^{\mathrm{p}}_{u,a}$ and requests in $R_\ver(I_\node)$. The red dot represents the start time of interval $I_\node$, black dots represent the arrivals of requests, the dashed curve illustrates the interval $I_\node$, and the solid curves illustrate matches made by $\OPT_\ver$. $\node$ is even and $\hl(\node)$ is $\node$-odd. We can let $g_\node(t)=p_i$ when $p_i$ is pending under $\OPT_\ver$, for $i=1,2$. In this figure, $\ql(\node)$ is $p_1$, $\Il_\node=[a_\ver(\node), a(p_1))$ and $\Ir_\node=[a(p_1),m_\ver(\node))$.} 
\end{figure}

\begin{lemma}\label{chargingalignednoh}
Let $\node$ be an aligned node. If $\node$ is left-preempted, then $z_{\node}\leq c(\hl(\node)) + \int_{\Jr_\node} \hat{f}'_{g_\node(t)}(t)dt$; otherwise, $z_{\node}\leq \int_{\Jr_\node} \hat{f}'_{g_\node(t)}(t)dt$.
\end{lemma}
\begin{proof}
We have 
\[z_\node 
= \int_{\Jl_\node}f'_{\ver,r(\node)}(t)dt + \int_{\Jr_\node}f'_{\ver,r(\node)}(t)dt = z_\node(a(\ql(\node))) + \int_{\Jr_\node}f'_{\ver,r(\node)}(t)dt.\] We now bound the two terms separately. According to \cref{odd counter,chargable}, $g_\node(t)$ is defined for every $t\in \Jr_\node$. The fact that $g_\node(t)\succeq r(\node)$ and the noncrossing property of the consolidated instance imply the following bound on the  integral: $\int_{\Jr_\node} f'_{\ver,r(\node)}(t) dt = \int_{\Jr_\node} \hat{f}'_{r(\node)}(t) dt\leq \int_{\Jr_\node} \hat{f}'_{g_\node(t)}(t) dt$. 

It remains to bound $z_\node(a(\ql(\node)))$. If $\node$ is not left-preempted, then $\node$ must be odd and so $z_\node(t)$ is frozen before the arrival of $\ql(\node)$, we are done. Otherwise, $\hl(\node)$ is $\node$-odd (or $\node$-even, resp.) and $\node$ is only active during even (or odd, resp.) subintervals if $\node$ is even (or odd, resp.), we have that $z_\node(a(\ql(\node))) = z_\node(a(\hl(\node)))$. The arrival time of  $\rho(\hl(\node))$ is earlier than $a_\ver(\node)$, and there are two cases: $\rho(\hl(\node))\preceq r(\node)$, or $r(\node)\prec \rho(\hl(\node))$. If $\rho(\hl(\node))\preceq r(\node)$, then
\[z_\node(a(\hl(\node))) \leq f_{\ver,r(\node)}(a(\hl(\node)))\leq\hat{f}_{r(\node)}(a(\hl(\node))) \leq \hat{f}_{\rho(\hl(\node))}(a(\hl(\node)))=c(\hl(\node)),
\]
according to the noncrossing property that the total delay of a later request is no more than that of an earlier one. If $r(\node)\prec \rho(\hl(\node))$, then
\begin{align*}
 z_\node(a(\hl(\node)))&\leq f_{\ver,r(\node)}(a(\hl(\node)))-f_{\ver,r(\node)}(a_\ver(\node))\\
&=\hat{f}_{r(\node)}(a(\hl(\node)))-\hat{f}_{r(\node)}(a_\ver(\node)) \\
&\leq \hat{f}_{\rho(\hl(\node))}(a(\hl(\node)))-\hat{f}_{\rho(\hl(\node))}(a_\ver(\node))\\
&\leq \hat{f}_{\rho(\hl(\node))}(a(\hl(\node)))=c(\hl(\node))
\end{align*}
according to the noncrossing property that the delay increase rate of an earlier request is no more than that of a later one. Because $z_\node(a(\ql(\node)))=z_\node(a(\hl(\node)))$, in both of the above cases it holds that
\[z_\node(a(\ql(\node)))\leq c(\hl(\node)).\]
This completes the proof of the lemma.
\end{proof}

We now use \cref{chargingalignednoh} to charge the total counter value of $\pr_{\ver,a}$ to $\OPT_\ver$ and prove the following. 

\begin{lemma}
    $\sum_{\node\in \Node^{\mathrm{p}}_{u,a}}z_{\node}\leq 2\OPT_\ver$.    
\end{lemma}

\begin{proof}
Consider the following charging scheme. For each node $\node \in \mathcal{N}^{\mathrm{p}}_{u,a}$,  we charge $\hat{f}'_{g_\node(t)}(t)$ to the request $g_\node(t)$ for each $t \in \Jr_\node$. Moreover, if $\node$ is left-preempted, then we also charge $c(\hl(\node))$ to the request $\hl(\node)$. We call the first type of charging \emph{instantaneous charging}, and the other \emph{non-instantaneous charging}. By \cref{chargingalignednoh}, the total charge is at least $\sum_{\node \in \mathcal{N}^{\mathrm{p}}_{u,a}}z_\node$. 

We now show that we do not overcharge $\OPT_\ver$. We first bound the total charge due to instantaneous charging. Since at most one node can be active at any time, at each time $t$, there is at most one request $r$ that is pending in $\OPT_\ver$ that is charged $\hat{f}'_{r}(t)$, and $\OPT_\ver$ incurs an instantaneous delay cost of $\hat{f}'_{r}(t)$. Thus, the total instantaneous charge is at most $\OPT_\ver$.

Every request $r$ also receives a non-instantaneous charge of $c(r)$ for each node $\node \in \mathcal{N}^{\mathrm{p}}_{u,a}$ for which $r = \hl(\node)$. Since no two nodes in $\mathcal{N}^{\mathrm{p}}_{u,a}$ are left-equivalent, we get that it is charged at most once. Moreover, only requests that arrive later in a matched pair by $\OPT_\ver$ can receive a non-instantaneous charge. Thus, the total non-instantaneous charge is at most $\sum_{(r,r') \in M^*_\ver} c(r') = \OPT_\ver$. 

In conclusion, the total charge is at least $\sum_{\node\in \mathcal{N}^{\mathrm{p}}_{\ver,a}} z_\node$, and is at most $2\OPT_\ver$. As a result, we have $\sum_{\node \in \Node^{\mathrm{p}}_{u,a}} z_\node \leq 2\OPT_\ver$, as desired.
\end{proof}

Combined with \cref{lem:prune}, we have the following corollary, which immediately implies \cref{aligned total}, and will be used in the following subsection to charge $\pr_{u,m}$ to $\OPT_\ver$.
\begin{corollary}
\label{all aligned}
$\sum_{\node\in\Node_{u,a}}z_\node\leq 8\OPT_\ver$.
\end{corollary}

\subsubsection{Charging misaligned $\pr_{u,m}$ to $\OPT_\ver$}
\label{sec:misalign-charging}
Next we consider requests in $\pr_{u,m}$. This subsection will be devoted to proving the following result:

\begin{lemma}[Bounding Misaligned Counters]\label{m total cost}
  $\sum_{\node\in\pr_{u,m}}z_{\node}\leq 28\OPT_\ver$.
\end{lemma}

\paragraph{Roadmap.} We first prove properties of $\OPT_\ver$ and $\ALG_\ver$ (\cref{even,donimation property}). Next, we define $\node$-crossing requests and split children, and use these properties to define an $\OPT_\ver$-induced canonical partition that we will charge $Y_\node$ to. Finally, we put these together to prove \cref{m total cost}.

In the previous section where we considered the case $\node\in \pr_{\ver,a}$, we are able to directly charge the instantaneous increase of $z_\node$ for each  $t \in \Jr_\node$ to a  request $g_\node(t)$ that is also pending under $\OPT_\ver$, while this fails for $\node\in\pr_{\ver,m}$. Therefore, we now need to better exploit the structure of $C_\node$ and the  property that $z_\node=2Y_\node$ (\cref{zr and cr,lemma: misaligned primary}), to charge $z_\node$ against $\OPT_\ver$ through $Y_\node$ as the medium.

The following lemma provides us with more insight into the relation between a node $\node\in\md_{u,m}$ and its children, which will be used when we charge the potential $Y_\node$ to $\OPT_\ver$.

\begin{lemma}\label{donimation property}
For every $\node\in\Node_\ver$ and $\node'\in C_\node$, if $a_\ver(\node')\in I^{\mathrm{odd}}_\node$, then $\node$ and $\node'$ have the same parity; if $a_\ver(\node')\in I^{\mathrm{even}}_\node$, then $\node$ and $\node'$ have opposite parities. Furthermore, for $\node\in\md_{u,m}$, $|R_\ver(I_{\node})|$ is odd (or even, resp.) if $\node$ is even (or odd, resp.). 
\end{lemma}
\begin{proof}
For every node $\node\in\Node_\ver$ and $\node'\in C_\node$, $r(\node)\prec r(\node')$ according to \cref{lemma:order on the chain}. We discuss two different cases: $a(r(\node'))$ is before $a_\ver(\node)$, $a(r(\node'))$ is after $a_\ver(\node)$.

In the first case, let $k_1$ denote the number of requests that arrive during $(a(r(\node)),a(r(\node')))$ and are unmatched under $\ALG_\ver$ at $a_\ver(\node)$, $k_2$ denote the number of requests that arrive during $(a(r(\node')),a_\ver(\node)]$ and are unmatched under $\ALG_\ver$ at $a_\ver(\node)$, $k_3$ denote the number of requests that arrive during $(a_\ver(\node),a_\ver(\node')]$ (note that $r(\node')$ is not included in any of the three parts). By definition, $\node$ is odd (or even, resp.) is equivalent to that $k_1+k_2+1$ is odd (or even, resp.). Because $\node'$ is a child of $\node$, all requests that arrive during $(a(r(\node)),a(r(\node')))$ must have been matched by time $a_\ver(\node')$ under $\ALG_\ver$. By \cref{lemma: interval match parity}, they can only be matched with requests that arrive later than $r(\node)$. As a result, at time $a_\ver(\node')$, the parity of the number of unmatched requests under $\ALG_\ver$ that arrive after $r(\node')$ is the same as $k_2+k_3-k_1$, which differs from $k_1+k_2+1$ if and only if $k_3$ is even, if and only if $a_\ver(\node')\in I^{\mathrm{even}}_\node$.

In the second case, let $k_1$ denote the number of requests that arrive during $(a(r(\node)),a_\ver(\node)]$ and are unmatched under $\ALG_\ver$ at $a_\ver(\node)$, $k_2$ denote the number of requests that arrive during $(a_\ver(\node),a(r(\node')))$, $k_3$ denote the number of requests that arrive during $(a(r(\node')),a_\ver(\node')]$. The parity of $\node$ has the same parity as $k_1$. At time $a_\ver(\node')$, all requests that arrive between $r(\node)$ and $r(\node')$ have been matched under $\ALG_\ver$. Similar to the previous case, by \cref{lemma: interval match parity}, the number of unmatched requests under $\ALG_\ver$ at $a_\ver(\node')$ that arrive after $r(\node')$ has the same parity as $k_3-(k_2+k_1)$, which differs from $k_1$ if and only if $k_2+k_3+1$ is even, if and only if $a_\ver(\node')\in I^{\mathrm{even}}_\node$.

Now we come to the last statement. By the definition of primary nodes and \cref{lemma: misaligned primary}, it holds that every misaligned primary node $\node\in\md_{u,m}$ is the critical parent when it is terminated, and so $z_\node=2Y_\node$ (\cref{zr and cr}). Let $\node'$ be the critical child, then $\node'$ is active at $m_\ver(\node)$, which implies that the number of unmatched requests under $\ALG_\ver$ that arrive after $r(\node')$ is even at $m_\ver(\node')=m_\ver(\node)$. Since no request that arrives between $r(\node)$ and $r(\node')$ is unmatched under $\ALG_\ver$ at the moment, the number of unmatched requests under $\ALG_\ver$ that arrive after $r(\node)$ is odd at $m_\ver(\node)$. If $\node$ is even (or odd, resp.), then this amount at $a_\ver(\node)$ is even (or odd, resp.), so the number of new arrivals during $I_\node$ (the only event that changes its parity), $|R_\ver(I_{\node})|$, is odd (or even, resp.).
\end{proof}

We now show how to construct the $\OPT_\ver$-induced canonical partition that we will charge $Y_\node$ to. For every $\node\in\pr_{u,m}$, we define a set of $\node$\emph{-crossing requests} based on how $\OPT_\ver$'s matches intersect $I_\node$, as well as \emph{split children} according to how those requests' matches of $\OPT_\ver$ intersect each of $\node$'s child interval. We will use the $\node$-crossing requests to define a canonical partition of $C_\node$ and charge its value, and thus $z_\node=2Y_\node$,  to the cost that $\OPT_\ver$ incurred on the $\node$-crossing requests and the counters of some aligned requests, which we have already bounded in \cref{all aligned}.

\begin{definition}
    [$\node$-crossing, split children]
    Let $\node$ be a node in $\Node_\ver$ and $q$ be a request in $\mathcal{R}_\ver$. We say that $q$ and its pending interval $I^*_q$ under $\OPT_\ver$ are \emph{$\node$-crossing} if $I^*_q$ is maximal among $\OPT_\ver$'s requests' intervals $I^*$ that intersect $I_\node$ but does not contain $I_\node$, i.e., $\emptyset \subsetneq I^* \cap I_\node \subsetneq I_\node$. Moreover, $I^*_q$ is \emph{internal $\node$-crossing} if, in addition, $I^*_q \subseteq I_\node$. The set of $\node$-crossing requests is denoted by $\cross_\node$.

 A child node $\node'\in C_\node$ is said to be \emph{split} by an $\node$-crossing request $q$ if it is the earliest request in $X_\node$ such that $\rho(q) \in R_\ver(I_{\node'})$ or $q\in R_\ver(I_{\node'})$. The set of split children nodes is denoted by $\splitr_\node$.
\end{definition}

\begin{figure}[ht]
\centering

\begin{tikzpicture}[scale=0.8]
\coordinate (p0) at (1.5,-0.5);
\coordinate (q0) at (2.5,0.5);
\fill[lightgray] (p0) rectangle (q0);
\coordinate (p1) at (3,-0.5);
\coordinate (q1) at (3.75,0.5);
\fill[lightgray] (p1) rectangle (q1);
\coordinate (p2) at (4,-0.5);
\coordinate (q2) at (5.5,0.5);
\fill[lightgray] (p2) rectangle (q2);
\coordinate (p3) at (6,-0.5);
\coordinate (q3) at (9.5,0.5);
\fill[lightgray] (p3) rectangle (q3);
\coordinate (p4) at (10,-0.5);
\coordinate (q4) at (11.5,0.5);
\fill[lightgray] (p4) rectangle (q4);
\coordinate (p5) at (12,-0.5);
\coordinate (q5) at (13.5,0.5);

\coordinate (p6) at (14,-0.5);
\coordinate (q6) at (15.5,0.5);
\fill[lightgray] (p6) rectangle (q6);
\coordinate (p7) at (16,-0.5);
\coordinate (q7) at (16.5,0.5);
\fill[lightgray] (p7) rectangle (q7);

\draw[thick, ->] (-1,0) -- (19,0);%

\fill (0,0) circle (2pt);
\node[below] at (0,-0.5) {$q_1$};

\fill[red] (1,0) circle (2pt);
\node[below] at (1,-0.5) {$\node$};

\fill[red] (1.5,0) circle (2pt);
\node[below] at (1.5,-0.5) {$\node_1$};

\fill[red] (3,0) circle (2pt);
\node[below] at (3,-0.5) {$\node_2$};

\fill[red] (4,0) circle (2pt);
\node[below] at (4,-0.5) {$\node_3$};

\fill[red] (6,0) circle (2pt);
\node[below] at (6,-0.5) {$\node_4$};
\fill[red] (10,0) circle (2pt);
\node[below] at (10,-0.5) {$\node_5$};
\fill (11.2,0) circle (2pt);
\fill (11.8,0) circle (2pt);
\fill[red] (14,0) circle (2pt);
\node[below] at (14,-0.5) {$\node_6$};
\fill[red] (16,0) circle (2pt);
\node[below] at (16,-0.5) {$\node_7$};

\node[below] at (5,-0.5) {$\rho(q_1)$};

\node[below] at (7,-0.5) {$q_2$};
\node[below] at (12.5,-0.5) {$\rho(q_2)$};
\node[below] at (13.4,-0.5) {$q_3$};

\fill (2,0) circle (2pt);
\fill (4.4,0) circle (2pt);
\fill (5,0) circle (2pt);
\fill (7,0) circle (2pt);
\fill (7.7,0) circle (2pt);
\fill (8.4,0) circle (2pt);
\fill (9,0) circle (2pt);
\fill (10.7,0) circle (2pt);
\fill (12.5,0) circle (2pt);
\fill (13.4,0) circle (2pt);
\fill (15,0) circle (2pt);
\fill (17,0) circle (2pt);
\fill (18,0) circle (2pt);
\node[below] at (18,-0.5) {$\rho(q_3)$};

\draw[thick] (0,0) to[out=45, in=135] (5,0);
\draw[thick] (7,0) to[out=45, in=135] (12.5,0);
\draw[thick] (13.4,0) to[out=45, in=135] (18,0);
\end{tikzpicture}
\caption{Illustration of $\node$-crossing requests with 7 children $\node_1, \ldots, \node_7$ and 3 $\node$-crossing requests $q_1,\ldots,q_3$. In this figure, the red dots represent the starts of node intervals, the black dots represent the arrivals of requests in $\mathcal{R}_\ver$, the arcs represent the $\OPT_\ver$ matching for the $\node$-crossing requests, and the gray boxes represent for each child $\node_i$, its interval $I_{\node_i} = [a_\ver(\node_i), m_\ver(\node_i))$. The requests $q_1$ and $\rho(q_3)$ are not in $R_\ver(I_\node)$. The split children are $\node_3$ (split by $q_1$) and $\node_4$ (split by $q_2$).} %
\label{fig:crossing}
\end{figure}

\begin{proposition}
    \label{prop:crossing}
    Let $\node \in \pr_{u,m}$, $\node_1 \prec_\ver \ldots \prec_\ver \node_k$ be its children, and $q_j$ be the $j$-th earliest $\node$-crossing request. Then, $q_1 =\rho(\hl(\node))$ if $\node$ is left-preempted, otherwise $q_1$ is the first request that arrives during $I_\node$. For each $j>1$, $q_j$ is the next request that arrives after $\rho(q_{j-1})$. Moreover, for each $j>1$, the request $q_j$ is $\node$-odd (or $\node$-even, resp.) if $\node$ is even (or odd, resp.). Finally, if $q$ is the latest $\node$-crossing request, then $q=\hr(\node)$.
\end{proposition}

\begin{proof} 
    
 If $\node$ is left-preempted, first of all, observe that $\rho(\hl(\node))\in X_\node$: otherwise if $I^*_{\rho(\hl(\node))}$ is not maximal, then there is a request in $R_\ver(I_\node)$ that arrives later than $\hl(\node)$ and is matched by $\OPT_\ver$ with another that arrives before $a_\ver(\node)$, a contradiction to the definition of $\hl(\node)$. Therefore, $q_1\preceq \rho(\hl(\node))$ by definition that $q_1$ is the earliest request in $X_\node$, which then implies that $\rho(q_1)\succeq \hl(\node)$ by laminarity. As a result, $\rho(q_1)$ must be $\hl(\node)$ and $q_1=\rho(\hl(\node))$. The case that $\node$ is not left-preempted is easy to show analogously.
    
    It is easy to see that $q_j\succ \rho(q_{j-1})$, otherwise if $q_{j-1}\prec q_j\prec \rho(q_{j-1})$, then $I^*_{q_j}$ is not maximal among $\OPT_\ver$ intervals that intersect $I_\node$ as it is contained in $I^*_{q_{j-1}}$. Suppose, towards a contradiction, that there is a request that arrives after $\rho(q_{j-1})$ and before $q_j$. Let $s$ be the next request that arrives after $\rho(q_{j-1})$. Observe that $s$ is not an $\node$-crossing request as it would have implied that $s = q_j$.
     By laminarity of $\OPT_\ver$, there are 3 possible cases: (1) $I^*_{\rho(s)}$  contains $I^*_{q_{j-1}}$; (2) $I^*_{s}$ contains $I^*_{q_j}$; or (3) $\rho(q_{j-1}) \prec \rho(s) \prec q_j$. The first two cases are ruled out by the fact that $q_j$ and $q_{j-1}$ are $\node$-crossing. The last case is also ruled out since it implies that $s$ should have been an $\node$-crossing request. Since every possibility leads to a contradiction, we get the first part of the proposition.

Moreover, since $\rho(q_1)$ is $\node$-even (or $\node$-odd, resp.) if $\node$ is even (or odd, resp.), $q_j$ is the next request that arrives after $\rho(q_{j-1})$, and the number of requests that arrive between $q_j$ and $\rho(q_j)$ is even for every $j$ (\cref{even}), we get that $q_j$ is $\node$-odd (or $\node$-even, resp.) if $\node$ is even (or odd, resp.), for every $j>1$.

Finally, we show that $q=\hr(\node)$, where $q$ is the latest $\node$-crossing request. Because $q$ is $\node$-odd (or $\node$-even, resp.) if $\node$ is even (or $\node$-odd, resp.), if $\rho(q)\in R_\ver(I_\node)$, then $\rho(q)$ must be $\node$-even (or odd, resp.) if $\node$ is even (or odd, resp.), and the next request that arrives after $\rho(q)$ should be in $R_\ver(I_\node)$ (\cref{donimation property}) and thus $X_\node$ as well, a contradiction. Therefore, $\rho(q)$ arrives after $m_\ver(\node)$, and so $\node$ is right-preempted. By the maximality of $I^*_q$ and the definition that $\hr(\node)$ is the earliest in $R_\ver(I_\node)$ that is matched by $\OPT_\ver$ with a request that arrives after $m_\ver(\node)$, $\hr(\node)$ must be $q$.
\end{proof}
\paragraph{$\OPT_\ver$-Induced Partition.}
Let $\node \in \pr_{u,m}$, $\node_1 \prec_\ver \ldots \prec_\ver \node_k$ be its children, $\node_{b(1)} \prec_\ver \ldots \prec_\ver \node_{b(\ell)}$ be the split children, and $q_{b(j)}$ be request in $X_\node$ that splits $\node_{b(j)}$ for each $1\leq j\leq \ell$.
The \emph{$\OPT_\ver$-induced partition} $\mPo_\node$ of $C_\node$ consists of:
\begin{itemize}
\item singleton $\{\node_{b(j)}\}$ for each split child, $1\leq j\leq \ell$;
\item the singleton $\{\node_k\}$ if not already included in the previous step;
\item among all remained children of $\node$: every \emph{maximal} contiguous set $\{\node_{i},\ldots,\node_j\}$ such that there exists $q\in X_\node$ that $a(q)<a_\ver(\node_i)$ and $a(\rho(q))\geq m_\ver(\node_j)$; if such an $\node$-crossing request $q$ that $a(q)<a_\ver(\node')$ and $a(\rho(q))\geq m_\ver(\node')$ does not exist for child $\node'$, then singleton $\{\node'\}$.
\end{itemize}
\cref{prop:crossing} implies that $\mPo_\node$ indeed partitions $C_\node$. In the example in \cref{fig:crossing}, the partition $\mPo_\node$ consists of $\{\node_1, \node_2\}, \{\node_3\}, \{\node_4\},\{\node_5\},\{\node_6\},\{\node_7\}$.

In the remainder of this subsection, we describe the charging scheme for every $\node\in\pr_{u,m}$ individually (\cref{lem:m-charging}), and then in the proof of \cref{m total cost}, use \cref{lem:prune} to show that there is no overcharging.

We first observe that in the last case, if a non-split child $\node'$ satisfies that such an $\node$-crossing request $q$ that $a(q)<a_\ver(\node')$ and $a(\rho(q))\geq m_\ver(\node')$ does not exist, then there is no new request arrival in $\mathcal{R}_\ver$ during the whole $I_{\node'}$. In this case, if the counter of $\node'$ increases to a positive value, then the child counter of $\node'$ cannot be active (because $\node'$ and its child have opposite parities in this case according to \cref{donimation property}) and so $\node'$ cannot be terminated as a critical parent, a contradiction. As a result, $z_{\node'}$ must be zero, and we omit the singleton $\{\node'\}$ in the remaining analysis.

\begin{lemma}
    \label{lem:m-charging}
    For $\node \in \pr_{u,m}$, we have $Y_\node \leq \sum_{q \in \cross_\node} 2c(q) + \sum_{\node' \in \splitr_\node} z_{\node'}$.
\end{lemma}

\begin{proof}
    Observe that $\mPo_\node$ is a canonical partition and thus, $Y_\node \leq \sum_{B_\node\in\mPo_\node}v_\node(B_\node)$. We now show that $\sum_{B_\node\in\mPo_\node}v_\node(B_\node) \leq \sum_{q \in \cross_\node} 2c(q) + \sum_{\node' \in \splitr_\node} z_{\node'}$.

We consider every part $\{\node_i,\ldots,\node_j\}$ such that: $q\in X_\node$, $q$ arrives before $a_\ver(\node_i)$ and $\rho(q)$ arrives after $m_\ver(\node_j)$. If $j>i$, then its value $v_\node(\{\node_i,\ldots,\node_j\})$ is no more than $f_{\ver,r(\node)}(m_\ver(\node_j))-f_{\ver,r(\node)}(a_\ver(\node_i))$. There are two cases: $q\preceq r(\node)$, or $r(\node)\prec q$. In the first case,
\[
f_{\ver,r(\node)}(m_\ver(\node_j))-f_{\ver,r(\node)}(a_\ver(\node_i))
\leq \hat{f}_{r(\node)}(m_\ver(\node_j))
\leq \hat{f}_q(m_\ver(\node_j))\leq c(q);
\]
in the second case,
\[
f_{\ver,r(\node)}(m_\ver(\node_j))-f_{\ver,r(\node)}(a_\ver(\node_i))
\leq \hat{f}_{q}(m_\ver(\node_j))-\hat{f}_{q}(a_\ver(\node_i))
\leq \hat{f}_{q}(m_\ver(\node_j))
\leq c(q).
\]
If $j=i$, then $v_\node(\{\node_i\})=z_{\node_i}\leq f_{\ver,r(\node_i)}(m_\ver(\node_i))-f_{\ver,r(\node_i)}(a_\ver(\node_i))$, and we can analogously discuss two cases $q\preceq r(\node_i)$ or $r(\node_i)\prec q$ to prove that $f_{\ver,r(\node_i)}(m_\ver(\node_i))-f_{\ver,r(\node_i)}(a_\ver(\node_i))\leq c(q)$ always holds. Moreover, because of the maximality in defining each part, we do not charge different parts to the same $\node$-crossing request $q$.

If $\node_k$ is a split child, we are already done. Otherwise let $q$ be the latest request in $X_\node$, then $q$ must arrive before $a_\ver(\node_k)$. Because $\OPT_\ver$ matches $q$ after $m_\ver(\node)=m_\ver(\node_k)$ (\cref{prop:crossing}),
we can similarly show that $v_\node(\{\node_k\})\leq \hat{f}_q(m_\ver(\node_k))\leq c(q)$.

Combining the above gives us the desired upper bound on $\sum_{B_\node\in\mPo_\node}v_\node(B_\node)$ and thus on $Y_\node$.
\end{proof}
\begin{proof}[Proof of \cref{m total cost}]
Consider the following charging scheme. For each node $\node \in \pr_{u,m}$, we charge $2c(q)$ to each $\node$-crossing request $q \in \cross_\node$ that is used to cover a non-split part of $\mathcal{Q}_\node$, and $z_{\node'}$ to each split child $\node' \in \splitr_\node$. We call the first type of charging a \emph{crossing charge} and the second a \emph{split charge}. \cref{lem:m-charging} and \cref{zr and cr} imply that the total charge is at least $\frac{1}{2}\sum_{\node \in \pr_{u,m}} z_\node$. 

We now argue that the total charge is at most $14\OPT_\ver$. Let $q$ be a request. First, we bound the total crossing charge $q$ receives. Observe that if $q$ is an internal $\node$-crossing request and $q$ receives crossing charge for some $B_\node\in\mPo_\node$, such a node $\node$ must be unique: otherwise if there are two such nodes $\node\prec_\ver \node'$, then $I^*_q\subseteq I_{\node'}\subseteq I_{\node_i}$ for some $\node_i\in C_\node$ due to that $q\in R_\ver(I_\node)$ and laminarity of intervals, but in this case $q$ cannot receive crossing charge for any $B_\node\in\mPo_\node$. As a result, every $q$ can receive crossing charge for some $B_\node\in\mPo_\node$ as an internal $\node$-crossing request at most once, left-preempts at most one node and right-preempts at most one node---thus the total crossing charge that all requests receive is at most $6\OPT_\ver$.

Next, we bound the total split charge. Let $\node'$ be a split child of $\node\in\pr_{u,m}$ and $q$ be the request in $X_\node$ that splits it. If $q\in R_\ver(I_{\node'})$, then $q$ must be the first request that arrives during $I_{\node'}$ and so $\node'$ is not left-preempted. If $\node$ is even, then $q$ is $\node$-odd (\cref{prop:crossing}) and so $\node'$ arrives during $I^{\mathrm{even}}_\node$, which implies that $\node'$ is odd (\cref{donimation property}). Therefore, $\node'$ is aligned. If $\rho(q)\in R_\ver(I_{\node'})$ and $\node$ is even, then $\rho(q)$ is $\node$-even. If $\node'$ arrives during $I^{\mathrm{even}}_\node$, then $\node'$ is odd and $\rho(q)$ is $\node'$-even, $\node'$ is aligned; if $\node'$ arrives during $I^{\mathrm{odd}}_\node$, then $\node'$ is even and $\rho(q)$ is $\node'$-odd, $\node'$ is aligned as well. Similar arguments also hold for the case that $\node$ is odd. In summary, $\node'$ is an aligned node. As a result, $\cup_{\node\in \pr_{u,m}}S_\node$ is a subset of $\mathcal{N}_{u,a}$. By \cref{all aligned} from the earlier subsection, we have the following upper bound on the total split charge:
\begin{equation*}
    \sum_{\node'\in \cup_{\node\in \pr_{u,m}}S_\node}z_{\node'}\leq \sum_{\node\in \mathcal{N}_{u,a}}z_\node\leq 8\OPT_\ver.
\end{equation*}

We conclude that the total of crossing charge and split charge is at most $14\OPT_\ver$. Since the total charge is also at least $\frac{1}{2}\sum_{\node \in \pr_{\ver,m}} z_\node$, we have $\sum_{\node \in \pr_{\ver,m}} z_\node\leq 28\OPT_\ver$.
\end{proof}

By combining \cref{pruned primary set,aligned total,m total cost}, we can compare the total cost of all $\ALG_\ver$'s against that of all $\OPT_\ver$'s:
\begin{lemma}
\label{charge alg to opt}
$\sum_{\ver \in V}\ALG_\ver\leq (1152H-576)\sum_{\ver \in V}\OPT_\ver$.
\end{lemma}

We now complete the proof of \cref{constant-competitive}.
By \cref{lemma: reduction to single,charge alg to opt}, we have
\begin{equation*}
    \ALG\leq 4\sum_{\ver\in V}\ALG_\ver
    \leq (4608H-2304)\sum_{\ver\in V}\OPT_\ver,
\end{equation*}
where the last part of the equalities is no more than 
\[(9216H-4608)\SOL_d+(4608H^2-2304H) \SOL_t\]
for any global solution $\SOL$, according to \cref{lemm: levels of SOL}. Together with the randomized embedding of general metrics into $2$-HSTs with height $O(\log\Delta)$ with distortion ratio $O(\log n)$ (\cref{lem: randomized embedding}) and \cref{lem: cost conversion between metrics}, there is an $(O(\log n\log\Delta), O(\log^2\Delta))$-competitive algorithm on general $n$-point metrics with aspect ratio $\Delta$. In the special case of the single location, there is only one vertex $\ver$ of which no descendant filters out any request, and so $\mathcal{I}_\ver$ is the same as the consolidated instance and the constant competitiveness is straightforward from \cref{charge alg to opt} applied to $H=1$. This concludes the proof of \cref{constant-competitive}.

\section{Concluding Remarks and Open Problems}\label{sec: conclusions}

In this paper, we presented the first polylogarithmic-competitive algorithm on general metrics as well as constant-competitive algorithm on the single location for online matching problem with non-identical delay functions.  Our algorithm operates in the non-clairvoyant setting and applies to arbitrary noncrossing delay functions, thereby simultaneously removing the assumptions of identical delay functions, concavity, and clairvoyance that underlie all previous constant- or polylogarithmic-competitive algorithms. Our result substantially broadens the class of delay functions for which competitive algorithms are known and narrows the gap between online matching with delays and many other online optimization problems with delays.

Several interesting questions remain open. The most immediate one is whether the noncrossing assumption can be relaxed. As we show in \cref{prop:noncrossing-necessary} (\cref{sec:noncrossing-necessary}), some restriction on the delay functions is necessary, since arbitrary delay functions do not admit a bounded competitive ratio even in the clairvoyant setting. It would therefore be interesting to characterize the largest natural class of delay functions that admits a constant-competitive algorithm.

Another natural direction is to remove the dependence of the competitive ratio on the aspect ratio of the metric space, as it can be unbounded as a function of the number of points. In particular, can one design an $O(\log n)$-competitive algorithm for general metric spaces? Existing algorithms for linear and concave delays rely on the height reduction technique to embed general metrics into weighted HSTs with height $O(\log n)$. However, weighted HSTs do not preserve the symmetry that all leaf descendants of a vertex are at the same distance from it, which breaks our charging scheme as the total delay when different leaves' requests are promoted to the same vertex can differ significantly. Moreover, our method to charge a request's delay increase in the odd state to some counter increase within the subtree (\cref{lemma: subtree charge}) is more complex than the previous ones while losing another height factor due to properties of virtual algorithms and matches. If one can achieve an $O(\log n)$-competitive algorithm for general metric spaces, then it would be a unifying framework that includes the previous results on uniform linear and concave delays as special cases.

\section*{AI Usage}
All algorithms and proofs were developed by the authors without the use of AI tools. We subsequently used Gemini DeepReview, the paper-review tool suggested by the conference, and addressed various issues identified in its report, particularly those concerning Section 5.

\printbibliography
\end{document}